\documentclass[12pt]{article}
\usepackage[T1]{fontenc}
\usepackage{amsmath,amssymb,mathrsfs}
\usepackage{paralist}
\usepackage{slashed}
\usepackage{graphicx}
\usepackage{subfigure}
\usepackage{color}
\usepackage{fullpage}
\usepackage{multirow}
\usepackage{hyperref}
\usepackage{amscd}
\usepackage{xcolor}
\usepackage{braket}
\usepackage{amsfonts}
\usepackage{tikz}
\usepackage{pgfplots} 
\usepackage{cite}
\usepackage{empheq}

\catcode`\@=11 
\def\numberbysection{\@addtoreset{equation}{section} 
        \def\theequation{\thesection.\arabic{equation}}} 

\def\be{\begin{equation}} 
\def\ee{\end{equation}} 
\def\ba{\begin{eqnarray}} 
\def\ea{\end{eqnarray}} 
\def\bali{\begin{align}}
\def\eali{\end{align}}
 
\def\ov{\overline}

\def\Z{\mathbb{Z}}

\def\nl{\nonumber \\}

\def\de{\partial} 
\def\wt{\widetilde} 
 
\def\Tr{{\rm Tr}} 
\def\dag{\dagger} 
 
\def\a{\alpha}

\def\G{\Gamma} 
\def\D{\Delta} 
\def\d{\delta} 
\def\e{\epsilon} 
\def\eps{\varepsilon}

\def\k{\kappa} 
\def\l{\lambda} 
\def\L{\Lambda}

\def\r{\rho} 
\def\s{\sigma} 
\def\S{\Sigma} 
\def\t{\tau} 
\def\f{\phi} 
\def\vf{\varphi}

\def\w{\omega} 
\def\W{\Omega} 
 
\def\th{\theta}

\def\u1{\widehat{U(1)}}
\def\su2{\widehat{SU(2)}_1}

\begin{document} 
 
\begin{titlepage} 
\begin{center} 
 
\vskip .6 in 
{\LARGE Bulk-Boundary Correspondence}
\medskip

{\LARGE in the Quantum Hall Effect} 
 
\vskip 0.2in 
Andrea CAPPELLI ${}^{(a)}$ and 
Lorenzo MAFFI ${}^{(a,b)}$\\ 
 \medskip

{\em ${}^{(a)}$ INFN, Sezione di Firenze}\\ 
{\em ${}^{(b)}$ Dipartimento di Fisica, Universit\`a di Firenze\\ 
Via G. Sansone 1, 50019 Sesto Fiorentino - Firenze, Italy} 
\end{center} 
\vskip .2 in 
\begin{abstract} 
We present a detailed microscopic study of edge excitations for
$n$ filled Landau levels. We show that the higher-level wavefunctions 
possess a non-trivial radial dependence that should be integrated over 
for properly defining the edge conformal field theory. 
This analysis let us clarify the role of
the electron orbital spin $s$ in the edge theory and to discuss its
universality, thus providing a further instance of the bulk-boundary
correspondence. We find that the values $s_i$ for each level, $i=1,\dots,n,$
parameterize a Casimir effect or chemical potential shift that could
be experimentally observed.
These results are generalized to fractional and hierarchical fillings
by exploiting the W-infinity symmetry of incompressible Hall fluids.
\end{abstract} 
 
%\pacs{73.43.Cd, 11.25.Hf, 73.23.Hk, 73.43.Jn} 
\vfill 
 
\end{titlepage} 
\pagenumbering{arabic} 
\numberbysection

%-1--------------------------------------------- 
 
\section{Introduction} 
The geometric response \cite{Avron} of quantum Hall states
\cite{qhe-books} has been extensively investigated in the recent
literature \cite{Read} \cite{Haldane}. The low-energy
effective action has been extended by coupling to a background
metric, leading to the Wen-Zee terms \cite{Wen-Zee}
\cite{Abanov1}. Other methods have also been developed, such as
explicit wavefunction constructions \cite{grav-bulk} \cite{Wiegmann},
hydrodynamic theory \cite{hydro}, the W-infinity symmetry \cite{CR},
and bi-metric theories \cite{bi-metric}.

As in any topological phase of matter, the responses have
a dual manifestation in the bulk and at the edge of the system, and their
interplay is called the bulk-boundary correspondence \cite{qhe-books}.
The best-known example is given by the electromagnetic response: 
the bulk Hall current compensates the chiral anomaly
at the edge, i.e. the non-conservation of the boundary charge, according to
the so-called anomaly inflow mechanism \cite{CDTZ} (also known as Laughlin
flux argument \cite{Laughlin}).
Since the anomaly is an exact and universal feature of the
conformal field theory of edge excitations \cite{cft} \cite{heat}, and is
related to topological invariant quantities, we can infer that 
the bulk Hall conductivity $\s_H$ is also universal and exact. 

The first geometric response of the edge theory is given by the
gravitational anomaly parameterized by the difference of conformal
central charges $c-\bar c$ of chiral and anti-chiral edge modes (in
the simplest case $\bar c=0$).  This anomaly describes the thermal
Hall current and thermal conductivity $\k_H$ \cite{heat}, recently observed
experimentally \cite{heat-exp}, and its value is similarly exact and 
universal.  The
bulk-boundary correspondence is more subtle in this case, since the
heat current flows on the edges and not in the bulk \cite{heat'}.  Other bulk
effects parameterized by $c$ have nevertheless been recently found,
using effective actions \cite{Abanov2}, Berry phase calculations 
\cite{grav-bulk} \cite{Wiegmann} and hydrodynamic
approaches  \cite{hydro}.

In this paper we analyze the second geometric response involving
the electron orbital spin $s$ (intrinsic angular momentum) and
describe the associated bulk-boundary correspondence.
Originally introduced in the Wen-Zee action,
the quantity $s$ parameterizes several interesting bulk phenomena: a shift
in the linear relation between the number of electrons and fluxes for  
compact spatial geometries \cite{Wen-Zee}, the response
of the Hall fluid to shear, i.e. the Hall viscosity \cite{Avron} \cite{Read}
 \cite{Haldane},
the spatial modulation of the Hall current \cite{Son} and, finally,
the fluctuation of the ground-state density near the edge \cite{edge-bump}.
Furthermore, the combination $c-12\nu s^2$ parameterizes the Berry phase of the
ground-state \cite{grav-bulk}.

Since $s$ is a coupling constant of an action
of Chern-Simons type, it is independent of local dynamics
and universal for translation invariant systems. Moreover
it is associated to a topological quantity in compact geometries.
However, it is not related to an anomaly of the edge theory, and
its physical meaning at the edge has so far remained unclear.

The recent work \cite{AGJ} has shown that in geometries with a
boundary, the Wen-Zee action is modified by the addition of two
boundary terms parameterized by the orbital spin, that are local edge
actions and thus non-anomalous.  This result implies that the Hall
fluid can have interfaces where the averages over electron species
$\ov s$ and $\ov{s^2}$ change values
without breaking any symmetry or closing the gap. In the disc
geometry, these local terms also predict ground-state values of the charge
and conformal spin at the edge.

In this paper, we describe the role of the orbital spin within edge physics
and explain the universal features that can be associated
to this quantity. We first build the theory of edge excitations for
$n$ filled Landau levels, by taking a straightforward limit of the
microscopic states near the edge.
This analysis was not done in the past beyond the lowest level $n=0$ 
\cite{CDTZ},
to the best of our knowledge, and is relevant for our problem, because
the orbital spin takes different values, 
$s_i=(2i+1)/2$ in the levels $i=0,1,\dots,n-1$.
We consider the geometry of the disc with radius $R$, realized by the
infinite plane with a confining potential; we find that the edge states are
localized at values of the radius $r\sim R$, 
and for $i>0$ rapidly oscillate in a region of the size of the magnetic 
length, $|r-R|=\Delta$ with $\Delta=O(\ell)$.
We then define the edge currents $\rho^{(i)}(\th)$ 
that are bilinears of Fermi
fields and obtain the chiral conformal field theory (Luttinger theory) with
$n$ independent branches and central charge $c=n$ \cite{qhe-books}.

Next we discuss the edge spectrum and show that the $s_i$ cause a shift
in the dispersion relation, up to a global constant, 
$s_i\to s_i +s_o$, that is  independent of bulk dynamics.
The $s_i$ can be included in the Hamiltonian of the conformal theory, 
i.e. the Virasoro generator $L_0$, by a redefinition of the chemical potential
$\mu_i\to \mu_i-s_i$ that becomes level dependent.

We describe some physical consequences of this fact in several
settings at the edge. In the general case of smooth boundary and
non-interacting branches of excitations, this effect is not
observable.  In the case of a disc with a sharp boundary, and for
interacting branches, we find non-vanishing ground-state values of the
edge charge and conformal spin that are parameterized by the differences
$s_i-s_j$, and are in agreement with the results of the effective action
approach \cite{AGJ}.  These are Casimir-like effects associated to
the orbital spin. 
Note that ground-state values of relativistic theories are not
observable in general, but relative differences do make sense.

In the second part of this work, we show how to generalize these results
to fractional fillings $\nu<1$ by using the W-infinity symmetry of
incompressible Hall fluids \cite{CTZ1} \cite{Sakita} \cite{CTZ2}.
Namely, we describe the edge excitations by deforming the density with
area-preserving diffeomorphisms of the plane.  We thus recover the
edge expansion at $r\sim R $ with its radial dependence
characterizing the different branches. In particular, the Jain
hierarchical states with $\nu=n/(qn+1)$, $q$ even, possess $n$ branches of
edge states with orbital spins $s_i=i+(q+1)/2$, $i=0,\dots,n-1$,
corresponding again to $s_{i}-s_{j}\in \Z $. These quantities parameterize
the ground-state values of edge charge and conformal spin.

In last section, we briefly discuss the possibility of measuring these
effects by a tunneling experiment in the Coulomb blockade regime
\cite{Coulomb} and by quadrupole deformation of the confining
potential \cite{quadrup}.

The outline of the paper is as follows. In section 2, we recall some
facts about the Wen-Zee action and the results of Ref.\cite{AGJ}
concerning the orbital spin at the boundary.  In section 3 we
perform the microscopic derivation of the edge conformal theory for
$n$ filled Landau levels. In section 4, we describe the physical
consequences of the orbital spin.  In section 5, we reobtain the
multicomponent edge theory by using the W-infinity transformations of
the bulk and extend the analysis of orbital spin effects to fractional
fillings.  In section 6, we discuss possible experiments.  Finally,
the conclusions present some remarks and open problems.  The two
appendices contain the analysis of non-relativistic corrections to
the conformal theory and the calculation of the Landau level 
spectrum with general boundary potential.

%-2--------------------------------------------- 
\section{Effective actions and geometric response}

In this section, we introduce the 
effective action of quantum Hall states, in the multicomponent case
that will be relevant for the following analysis \cite{qhe-books}.
The geometric part of the action, not involving any local dynamics,
can be derived by assuming that the low-energy fluctuations 
of the incompressible fluid are described by conserved currents
$j_{(i)}^\mu$, that are dual to hydrodynamic gauge fields $a_{(i)\nu}$,
$\mu,\nu=0,1,2$, as follows:
\be
j_{(i)}^\mu =\frac{1}{2\pi}\eps^{\mu\nu\r}\de_\nu a_{(i)\r},
\qquad i=0,\dots,n-1,
\ee
for any component $(i)$.
The gauge fields interact by a Chern-Simons topological action,
with couplings specified by the so-called $K$ matrix.
The system is placed in an electromagnetic background $A_\mu$ and
a (time-dependent) spatial metric $g_{ij}$, $i,j=1,2$, whose associated
spin connection is Abelian, $\w_\mu=\eps_{ab}\w_\mu^{ab}/2$, where
$a,b=1,2$ are local frame indices \cite{grav-bulk}.
Using the notation of forms, i.e. $a=a_\mu dx^\mu$, and summing over the
$n$ fluid components, we write the following effective action:
\be
S[a,A,\w]=-\frac{1}{4\pi}\int K_{(i)(j)} a_{(i)}da_{(j)}
+\frac{1}{2\pi}\int a_{(i)} \left(
t_{(i)} dA +s_{(i)} d\w \right),
\ee
where $t_{(i)}$ are the charges of fluid components (conventionally,
$t_{(i)}=1$ $\forall i$) and $s_{(i)}$ are their orbital spin values.

The integration over the $a_{(i)}$ field yields the induced action 
\cite{Abanov1},
\be
S_{\rm ind}[A,g] = \frac{\nu}{4\pi}\int
\left(AdA\ +\ 2 \ov{s}\; \w dA\ +\ \ov{s^2}\; \w d\w \right)\ +\
\frac{c}{96\pi}\int \Tr\left(
\G d\G + \frac{2}{3}\G^3 \right),
\label{i-act}
\ee
with the following expressions for the parameters \cite{qhe-books}:
\be
\nu=t_{(i)}K^{-1}_{(i)(j)} t_{(j)}, \qquad
\nu \ov{s} =s_{(i)}K^{-1}_{(i)(j)} t_{(j)}, \qquad
\nu \ov{s^2} =s_{(i)}K^{-1}_{(i)(j)} s_{(j)},\qquad c=n,
\label{nusbars}
\ee
corresponding to the filling fraction $\nu$, 
the average orbital spin $\ov{s}$, the average square $\ov{s^2}$
and the central charge $c$, respectively. The conventional naming 
for these quantities 
refers to the case of integer filling, where $K$ is the identity 
matrix (we only discuss the case of positive definite $K$, for simplicity).
The $s_{(i)}$ values  for $\nu=n$ can be computed from the total 
angular momentum $M$ of each level filled with $N$ electrons, 
using the formula:
\be
M=\frac{N^2}{2\nu}-N s,
\label{tot-moment}
\ee
where $\nu=1$ for each level. One finds:
\be
s_{(i)}=\frac{2i+1}{2}, \qquad i=0,\dots,n-1.
\label{si}
\ee

In the expression of the induced action (\ref{i-act}), 
the first term is the Chern-Simons form $S_{CS}$ responsible for the
Hall conductance, and the second and third terms are called
Wen-Zee  $S_{WZ}$ and gravitational Wen-Zee terms $S_{gWZ}$, respectively.
The last term, the gravitational Chern-Simons term $S_{gCS}$, 
involves the Christoffel connection  $(\G)^\mu_\nu=\G^\mu_{\nu\l}dx^\l$
and arises from the measure of integration
for the $a_{(i)}$ path-integral, owing to the framing anomaly 
\cite{Abanov-framing}.

In a spatial geometry $\S$ with a boundary, like the disk $D_2$, the action
(\ref{i-act}) is invariant under gauge transformations
and diffeomorphisms up to total derivatives that
must be compensated by additional boundary terms.
The solution of this problem is well-know and leads to the existence of
massless edge excitations described by a conformal field
theory defined on the spacetime boundary of the cylinder 
$\de {\cal M}=S_1\times\mathbb{R}$ \cite{qhe-books}.
In the present discussion, this is given by
the so-called Abelian theory of $n$ chiral massless bosonic fields $\vf_{(i)}$
(chiral Luttinger theory), with action $S_{CFT}[\vf_{(i)},A,g]$, also
involving the coupling to the $A$ and $g$ backgrounds. 
It can be shown that the variation of $S_{CFT}$ under gauge transformations
and diffeomorphisms cancels the corresponding variations of the two
terms $S_{CS}$ and  $S_{gCS}$ in the bulk action, respectively. 

The conformal theory at the edge is characterized by chiral
and gravitational anomalies; the first one corresponds to the
non-conservation of the edge charge and matches the bulk
Hall current by the anomaly inflow mechanism \cite{CDTZ}.
The gravitational anomaly amounts to non-conservation of energy-momentum
in the cylinder and leads to the heat current.
These anomalies cannot be eliminated by adding
terms in the action that are local in two dimensions, the spacetime where
the conformal theory is defined. Actually they are compensated by
actions that are local in one extra dimension, $S_{CS}$ and  $S_{gCS}$,
that are thus uniquely determined.
Furthermore, the anomalies are exact quantities in the conformal theory 
that are independent of short-distance physics \cite{cft}. These results
ensure that the coefficients $\nu$ and $c$, respectively parameterizing the
electric and thermal conductance $\s_H$ and $\k_H$, are
universal quantities independent of the details of the system. 
This is the bulk-boundary correspondence associated to these responses.

Next, we discuss the boundary terms needed to correct the non-invariances of 
the second and third terms, $S_{WZ}$ and $S_{GRWZ}$ in the action (\ref{i-act}). 
They
have been recently found in Ref.\cite{AGJ} and read:
\be
S_{WZ,b}=\frac{\nu\ov{s}}{2\pi}\int_{\de {\cal M}} A K, \qquad 
S_{gWZ,b}=\frac{\nu\ov{s^2}}{4\pi}\int_{\de {\cal M}} \w K,
\label{b-act}
\ee
where the integrals are taken on the boundary cylinder,
$A$ and $\w$ are the restrictions of the
one-forms on the boundary and $K$ is the one-form of the
extrinsic curvature of the boundary, $K=K_\a dx^\a$, $\a=0,1$.

The expressions (\ref{b-act}) are local in $(1+1)$ dimensions and thus
are not related to an anomaly of the edge theory. In generic quantum
field theories, local terms of the action are not universal in the
sense that they depend on the dynamics and can be modified, according
to different (possible) definitions of the renormalized quantities.
Actually, the local terms (\ref{b-act}) can also be considered for an
interface between two regions where $\ov{s}$ and $\ov{s^2}$ take
different values in the bulk \cite{AGJ}.  At this point the gap does
not vanish and there are no edge excitations.  In conclusions, the
results \eqref{b-act} suggest that the average orbital spin and its
square average are not universal quantities on the edge as well as in the
bulk in absence of translation invariance.

Summarizing, the effective action for the disk geometry is given by
the Eqs.(\ref{i-act}, \ref{b-act}):
\be
S_{\rm ind}[A,g] = S_{CS}+S_{WZ}+S_{WZ,b}+S_{gWZ}+S_{gWZ,b}+S_{gCS}+S_{CFT}.
\ee
The induced currents and response coefficients are found by 
taking variations of this action, as reviewed e.g. in Ref.\cite{CR}. 
Let us mention two results that are useful for the following analysis:
\begin{itemize}
\item
The total number of particles is given by the space integral of the
density, $\sqrt{g}\r_0=\d S_{\rm ind}/\d A_0$, and reads:
\ba
N&=&\frac{\nu}{2\pi}\int_\S \sqrt{g} B\ +\
\frac{\nu\ov{s}}{4\pi}\int_\S \sqrt{g}{\cal R}\ +\ 
\frac{\nu\ov{s}}{2\pi}\int_{\de \S}k + Q_{CFT}
\nl
&=& \nu N_\f +\nu\ov{s}\chi + Q_{CFT}.
\label{deg-flux}
\ea
In this expression, there appear the scalar curvature,
${\cal R}=2\eps^{ij}\de_i \w_j/\sqrt{g}$, the number $N_\f$ of 
magnetic fluxes through the surface and the Euler characteristic 
$\chi$.
Note that the bulk and boundary terms $S_{WZ}$ and $S_{WZ,b}$ in the action 
combine themselves to give the correct expression of the Gauss-Bonnet 
theorem for surfaces with a boundary, including the geodesic curvature
$k$, leading to $\chi=2-2h+b$, where $h$ and $b$ are the number of handles and boundaries, respectively.
\item
The spin density can be obtained by the variation of the action 
with respect to the spin connection at fixed metric,
$\sqrt{g}s^0=\d S_{\rm ind}/\d \w_0\vert_g$. There is an ambiguity
in performing this derivative, because metric and spin connection are
independent variables only in presence of torsion, that was not fully
accounted for in the derivation of (\ref{i-act}). 
At any rate, this ambiguity amounts to
trading angular momentum for spin. We are interested in the boundary 
contribution that originates from the term $S_{gWZ,b}$ (\ref{b-act}) for the
geometry of the flat disk $\left(\chi=1\right)$. It reads:
\be
{\cal S}_b=\int_{S_{1}} s^0 
= \frac{\nu\ov{s^2}}{4\pi}\int_0^{2\pi R}\! dx\; k
=\frac{\nu\ov{s^2}}{2}.
\label{t-spin}
\ee
\end{itemize}

Summarizing, in the geometry of the disk, the Wen-Zee action supplemented
by the boundary terms (\ref{b-act}) predicts non-vanishing ground-state values
for the spin (\ref{t-spin}) and the charge,
\be
Q_b = \nu\ov{s}+Q_{CFT},
\label{t-charge}
\ee 
at the boundary. In section 4, we shall recover and extend
these results by studying the edge conformal theory 
and explain to which extent these quantities are universal.

%-3----------------------------------------------
\section{Multicomponent edge theory}

In this section we find the conformal theory of edge excitations 
by taking a suitable limit of
the microscopic states of $n$ filled Landau levels.  This result will
set the stage for the analyses in this paper.  Although the limit to
the edge is rather straightforward, it was not yet analyzed in the
literature, beside the one-component case of the lowest Landau level
of Ref.\cite{CDTZ} and the nice work \cite{Dunne}.

We consider the Landau levels in the infinite plane with symmetric gauge
as discussed in Ref.\cite{CDTZ} and adopt the notations of that paper. 
In particular, the
magnetic length $\ell=\sqrt{2\hbar c/eB}$, $c$, $e$, $\hbar$ and the
electron mass $m$ are set to one, corresponding to $B=2$.
The Hamiltonian and angular momentum take the form:
\be
H= 2 a^\dag a +1,\qquad J=b^\dag b-a^\dag a,
\label{landau-ham}
\ee
in terms of two pairs of mutually commuting creation-annihilation operators,
\ba
a=\frac{z}{2}+ \ov{\de}, \qquad a^\dag=\frac{\ov{z}}{2} -\de, \qquad
\left[a,a^\dag\right]=1,
\nl
b=\frac{\ov{z}}{2}+ \de, \qquad b^\dag=\frac{z}{2} -\ov{\de}, \qquad
\left[b,b^\dag\right]=1,
\label{ab-oper}
\ea
involving the complex coordinate of the plane, $z=r\exp(i\th)$.
The single particle wavefunctions $\psi_{n,m}(z, \ov{z})$
are characterized by the values of the level index $n=0,1,\dots$ 
and angular momentum $m=-n,-n+1,\dots$. Their expression is:
\begin{equation}
\label{Landau-wfunct}
\begin{aligned}
\psi_{n,m}&=\frac{\left(b^{\dagger}\right)^{n+m}}{\sqrt{\left(n+m\right)!}}
\frac{\left(a^{\dagger}\right)^{n}}{\sqrt{n!}}
\frac{1}{\sqrt{\pi}}e^{-|z|^2/2}\\
&=\sqrt{\frac{n!}{\pi\left(n+m\right)!}}
z^{m}L_{n}^{m}\left(|z|^{2}\right)e^{-|z|^{2}/2},
\qquad m+n\geq 0,
\end{aligned}
\end{equation}
where $L_n^m$ are the generalized Laguerre polynomials.

%-3.1-----------------------------------------------------
\subsection{Limit of wavefunctions to the edge}

We consider the Hall state made by filling up to $N$ electrons per
level, thus forming a droplet of fluid of radius $R^2\sim N$, due to
the flux/degeneracy relation.  A confining radial potential
breaks the degeneracy of Landau levels near the boundary and create a
Fermi surface around that point. The specific form of the potential
will be discussed later and is not relevant momentarily.

The edge theory is defined by the states that describe particle-hole
excitations around the Fermi surface in a finite range of energy, i.e.
of angular momentum, in the limit $R\to\infty$.
The states in the $i$-th Landau level, $i=0,1,\dots$, filled with $N$ electrons
have momenta $-i\le m < N-i$. The edge theory 
is defined in the range \cite{CDTZ}:
\be
L-\sqrt{L} < m < L+\sqrt{L}, \qquad L\equiv R^2 \to\infty,
\label{ang-limit}
\ee
where $L\sim N$ is a given common momentum near the Fermi surface and the
range of $m$ is chosen to fit a linear spectrum of edge energies,
$\eps_m\sim v(m-L)/R $, i.e $\eps(k)=vk$ with $k$ the one-dimensional momentum.

\begin{figure}
\centering
\subfigure[\label{fig.1a}]{
\includegraphics[scale=0.9]{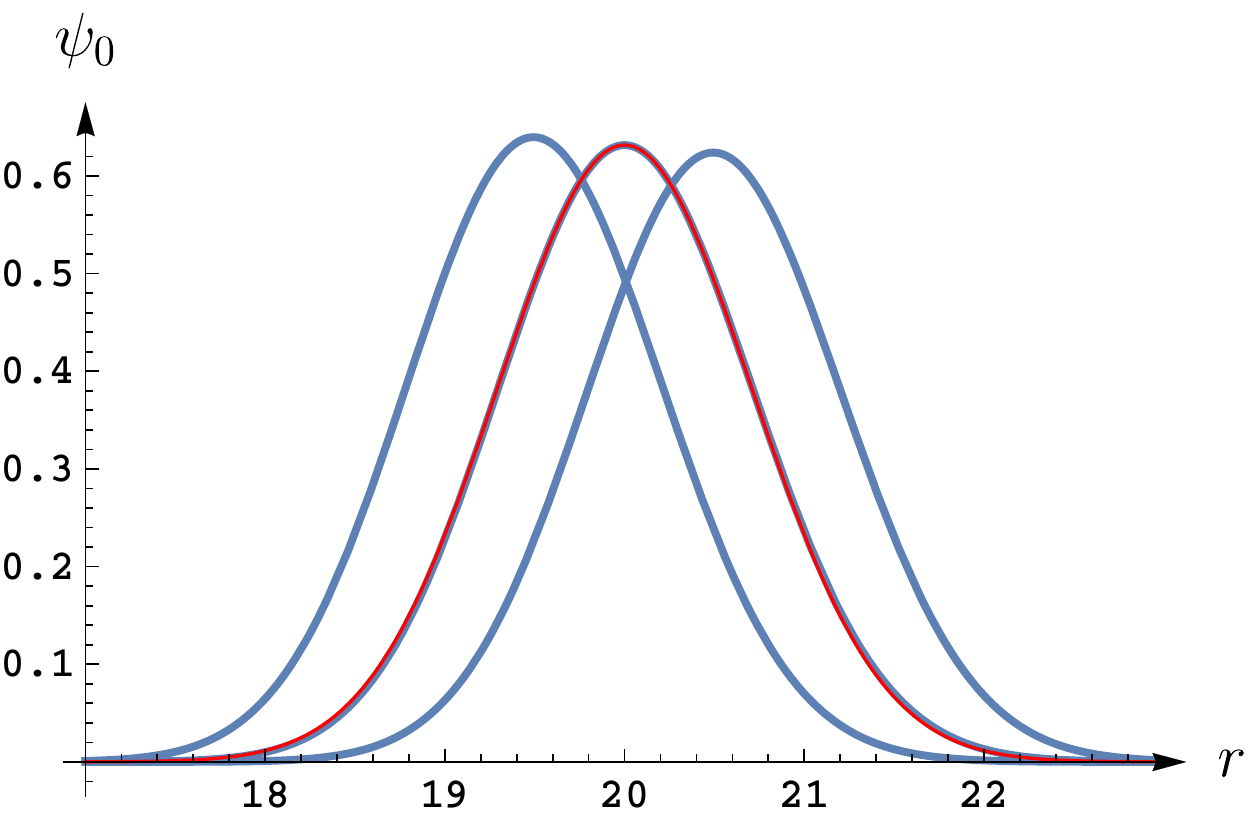}} 
\subfigure[\label{fig.1b}]{
\includegraphics[scale=0.9]{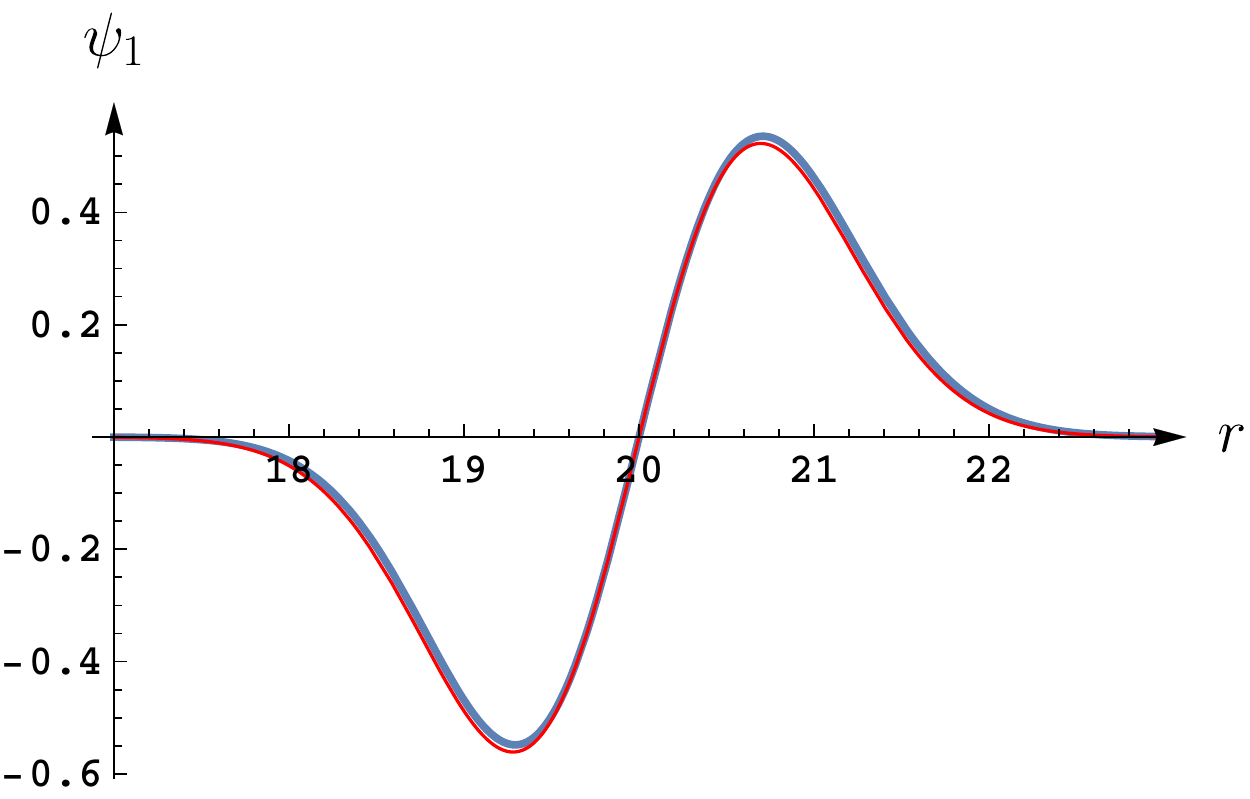}} 
\caption{(a): radial dependence of $\psi_{0,L+m'}$, for $L=R^{2}=400$
  and $m'=\lbrace -20,0,20\rbrace $ (blue), and its leading
  approximation (\ref{1LL-edge}) (red), for $m'=0$. (b): radial
  dependence of $\psi_{1,L-1+m'}$ (blue) and its approximation
  (\ref{2LL-edge}) (red), for $L=R^{2}=400$ and $m'=0$.}
\label{fig.1}
\end{figure}

As is well known, the wavefunctions $|\psi_{n,m}(r)|$ are localized around
the semiclassical orbits with $r^2\sim m$. Thus,
we can also expand the edge states for $r\sim R$ and
consider the combined limit for the angular momentum (\ref{ang-limit}) 
and the radial coordinate,
\be
r=R+x, \qquad x=O(1), \qquad R\to\infty. 
\ee
Let us perform this limit on the functions of the first level $n=0$:
we redefine the momentum w.r.t. the Fermi 
surface $m=L+m'$ and use the Stirling expansion. We obtain:
\ba
\label{1LL-edge}
\psi_{0,L+m'}\left(r,\theta\right)&\simeq& {\cal N}
\frac{e^{i\left(L+m'\right)\theta}}{\sqrt{2\pi R}}\ 
e^{-\left(x-\frac{m'}{2R}\right)^{2}}
\left(1+O\left(\frac{1}{R},\frac{m'}{R^{2}}\right)\right),
\nl
r&=&R+x,\qquad x=O(1), \qquad |m'|\le R, \qquad R^{2}=L\to\infty,
\ea
where the normalization constant is ${\cal N}=(2/\pi)^{1/4}$.
These wavefunctions are plotted in Fig.\ref{fig.1a}.

In the expression (\ref{1LL-edge}), the first factor corresponds 
to the wavefunction $\psi_{m'}(\th)=e^{im'\th}/\sqrt{2\pi R}$ for 
the $(1+1)$-dimensional Weyl fermion of the edge theory,
while the radial part is peaked at $r\sim R$ with spread $O(\ell)$ for
$R\to\infty$. In the earlier work \cite{CDTZ}, the states of the
edge theory were thus identified with the wavefunctions at 
fixed  radius $x=0$,
the remaining cut-off in momentum $\exp(-(m'/2R)^2)$ being irrelevant
for $R\to\infty$. 

It turns out that neglecting the radial dependence is not appropriate for
higher Landau levels. In order to understand the problem, let us consider 
the form of the second level wavefunctions:
\begin{equation}
\psi_{1,m}=\frac{z^{m}}{\sqrt{\pi\left(m+1\right)!}}\left(r^{2}-m-1\right)
e^{-r^{2}/2}.
\end{equation}
Upon taking the limit to the edge, we find: 
\begin{equation}
\label{2LL-edge}
\psi_{1,L+m'-1}\simeq
{\cal N}\frac{e^{i\left(L+m'-1\right)\theta}}{\sqrt{2\pi R}}\ 
2\left(x-\frac{m'}{2R}\right)
e^{-\left(x-\frac{m'}{2R}\right)^{2}}.
\end{equation}
The radial function now shows an oscillation of size 
$x=O(1)$ near the edge,  as shown in Fig.\ref{fig.1b}.

The oscillating behaviour is also present in the higher levels,
for the simple reason that the wavefunctions should be orthogonal
among themselves at fixed angular momentum, i.e. in $r$ space.
Owing to the Gaussian factor, it is rather easy to guess that
the radial functions should map into the harmonic oscillator basis involving
the Hermite polynomials $H_n$. This is indeed the case: 
in Eq.(\ref{2LL-edge}), the 
polynomial is identified as $(x-m'/(2R))\sim H_1(x-m'/(2R))$. The
inspection of the next few cases leads to the following result:
\be
\psi_{n,L+m}\left(R+x,\theta\right) \simeq
{\cal M} \frac{e^{i\left(m+L\right)\theta}}{\sqrt{2\pi R}}
H_{n}\left[\sqrt{2}\left(x-\frac{m+n}{2R}\right)\right]
e^{-\left(x-\frac{m+n}{2R}\right)^{2}},
\label{nLL-edge}
\ee
where ${\cal M}$ is another normalization factor.

The shift of the radial coordinate by $m/2R$ in (\ref{nLL-edge}) 
is also easily explained. In the limit $R\to\infty$,
the disk geometry can be approximated by the half plane, defined by $x<0$
in $(x,y)$ coordinates. In this geometry, the Landau levels in the linear
gauge $(A_x,A_y)=(0,Bx)=(0,2x)$ have the form:
\be
\wt\psi_{n,k_y}\left(x,y\right)\sim
e^{ik_yy}H_{n}\left[\sqrt{2}\left(x-\frac{k_{y}}{2}\right)\right]
e^{-\left(x-\frac{k_{y}}{2}\right)^{2}}.
\label{cil-wf}
\end{equation}
Owing to the periodicity of the edge of the disk, the half plane is
wrapped in the $y$ direction to form a cylinder, such that the
corresponding momentum is quantized by $k_y=m/R$. We thus recover
the expression (\ref{nLL-edge}) up to an overall phase for the
different gauge choice.

Note, however, that the functions \eqref{cil-wf} for the cylinder match
those of the disk \eqref{nLL-edge} with a level-dependent shift of
momentum $m/R\rightarrow\left(m+n\right)/R$ that will be crucial in
the following discussion. 
This difference is due to the extrinsic curvature of the boundary of the
disc, that vanishes for the half-plane and cylinder geometries. 
This is a first instance of the orbital spin $s_n$
in the edge theory.

In conclusion, the wave functions of edge states are Gaussians
localized in a spatial region $O(\ell)$ around $r=R$ and a momentum
range $O(\sqrt{L})$ around $L=R^2$. The spatial separation $\Delta
x=1/2R$ between neighbor states is in agreement with the
degeneracy/flux relation, since $N_{\phi}\rightarrow N_{\phi}+1$ amounts to
$R^2\rightarrow\left(R+1/2R\right)^2\sim R^2+1$.
The states of higher Landau level display an oscillating radial
dependence that is required by orthogonality.  
A glimpse of this fact had already appeared in analysis of
density shapes of Ref.\cite{Dunne}.

%-3.2-----------------------------------------------

\subsection{Multicomponent conformal theory}

We now construct the conformal theory of Weyl
fermions that describes the edge excitations for $\nu=n$.
The strategy is to consider bilinears of Fermi fields that are
observable quantities, in particular the modes of the density
$\hat{\rho}_{k}$ that are the building blocks of the Abelian
conformal theory \cite{cft}.

Let us start from the second-quantized field operator for $n$ Landau levels,
\begin{equation}
\label{field-op}
\hat{\Psi}\left(z,\bar{z}\right)=\sum_{m=0}^{\infty}\psi_{0,m}\hat{c}^{(0)}_{m}
+\sum_{m=-1}^{\infty}\psi_{1,m}\hat{c}_{m}^{(1)}+
\ldots+\sum_{m=-n+1}^{\infty}\psi_{n-1,m}\hat{c}_{m}^{(n-1)},
\end{equation}  
where $\hat{c}_m^{(i)}$ are fermionic destruction operator.
We consider the density,
$\hat{\rho}\left(z,\bar{z}\right)=\hat{\Psi}^{\dagger}\hat{\Psi}$,
and analyze the Fourier modes at the edge, also integrating over the
radial coordinate:
\begin{equation}
\label{den-modes}
\hat{\rho}_{k}\equiv\int_{0}^{\infty}dr \,r \int_{0}^{2\pi} 
d\theta\, \hat{\rho}\left(r,\theta\right)e^{-ik\theta}.
\end{equation}
This expression is analyzed in
the limit to the edge of the previous section, namely $R\to\infty$ with
$r=R+x$, $x=O(1)$, 
$R^{2}=L$, $m= L+m'$, $ |m'| < \sqrt{L}$.
First we substitute the field expansion (\ref{field-op}):
\begin{equation}
\begin{aligned}
\hat{\rho}_{k}=\int_{-R}^{\infty}\left(R+x\right)dx\,\int_{0}^{2\pi}& 
d\theta\sum_{m,n} 
\left(\psi_{0,L+m}\left(x,\theta\right)\hat{d}_{m}^{(0)}+
\psi_{1,L+m}\left(x,\theta\right)\hat{d}_{m}^{(1)}+\ldots\right)\times
\\
&\left(\psi_{0,L+n}^{*}\left(x,\theta\right)\hat{d}_n^{(0)\dagger}+
\psi_{1,L+n}^{*}\left(x,\theta\right)\hat{d}_{n}^{(1)\dagger}+\ldots\right)
e^{-ik\theta},
\end{aligned}
\end{equation} 
having redefined $\hat{d}^{(i)}_{m}\equiv\hat{c}^{(i)}_{L+m}$.
We then take the edge limit on wavefunctions and use
the orthogonality of Hermite polynomials; to
leading order $O(1)$ in the $1/R$ expansion, we find:
\begin{equation}
\label{cft}
\begin{aligned}
\hat{\rho}_{k}&=\sum_{m\in\Z}\left(\hat{d}^{(0)\dagger}_{m-k}\hat{d}^{(0)}_{m}+
\hat{d}^{(1)\dagger}_{m-k}\hat{d}^{(1)}_{m}+
\hat{d}^{(2)\dagger}_{m-k}\hat{d}^{(2)}_{m}+\ldots\right)+
\mathcal{O}\left(\frac{1}{R},\frac{k}{R}\right)
\\
&=\hat{\rho}_{k}^{CFT(0)}+\hat{\rho}_{k}^{CFT(1)}+\hat{\rho}_{k}^{CFT(2)}+
\ldots+\mathcal{O}\left(\frac{1}{R},\frac{k}{R}\right).
\end{aligned}
\end{equation}
In this calculation we neglected the shifts in the coordinates,
$x-\Delta m/R\sim x$, corresponding to the underlying 
low-momentum expansion.

The result (\ref{cft}) shows that the edge Fourier modes of the two-dimensional 
charge decompose into $n$
independent contributions $\hat{\rho}^{CFT(i)}_{k}$, $i=0,\ldots,n-1$, 
that act on the Fock spaces of the respective Landau levels.
Their expressions match the fermionic representation of 
the multicomponent Abelian conformal theory with $c=n$.
This result also agrees with the effective theory approach of 
section 2, where the edge fields $\vf^{(i)}$ realize the bosonic 
representation of the same conformal theory, i.e.  
$\hat{\r}^{CFT(i)}(\th)= \de_\theta\vf^{(i)}(\theta)$ \cite{qhe-books}.

Therefore, the $n$ edge densities are obtained by radial
integration of the  bulk density in the limit $R\to\infty$.
More precisely, the edge states  are found
by averaging the radial dependence of microscopic states 
near the edge in a shell $R-\D <r<R+\D$, with $\D= O(1)$, i.e. of the size of 
the ultraviolet cutoff. 
Other observables of the conformal theory are similarly obtained 
by radial integration of bilinear bulk quantities.
For example, the edge correlator is defined by:
\be
\label{edge-corr}
\langle\psi(\th)\psi^\dag(\th')\rangle_{CFT}\simeq
\int dr\; r\; \bra{\W}\Psi(r,\th)\Psi^\dag(r,\th')\ket{\W} .
\ee
This expression also decomposes in independent contributions 
for each branch of edge excitations.

In conclusion, the $n$ edge densities are obtained by radial
integration of the  bulk density in the limit $R\to\infty$.
More precisely, the edge states  are found
by averaging the radial structure of microscopic states 
near the edge in a shell $r-R= O(1)$, i.e. of the size of 
the ultraviolet cutoff.
Note that no non-locality is introduced by this approach, that
is simply a low-energy expansion.

Next, we observe that the contribution of one level
can be single out from the total current (\ref{den-modes}) by integrating
in $r$ with a suitable weight function $f_{j}\left(r\right)$, and
using the orthogonality of Hermite polynomials.
For example, let us suppose that we would like to remove the
contribution  of the lowest level $\hat{\rho}^{(0)}_0$ from the sum (\ref{cft}).
We can use the weight, 
\be
f_{(0)}\left(x\right)=1-4x^{2},
\ee 
and compute $\hat{\r}_k' =\int d^2x f_{(0)}\hat{\r}(r,\th)e^{-ik\theta}$.
We find that the lowest level does not contribute to leading order 
$O(1)$, because:
\begin{equation}
\int_{-\infty}^{\infty} drr\,f_{(0)}\,
\varphi_{0,L+m}\,\varphi_{0,L+m+k}^{*}=O\left(\frac{1}{R}\right).
\label{singleout}
\end{equation}

The fact that the Abelian currents appear at leading order $O(1)$ 
in the $R\to\infty$ expansion is
expected for conformal theories on the spacetime cylinder \cite{cft}. Higher
orders $O(1/R^k)$ correspond to conserved currents of higher
spin/conformal dimension $h=k+1$.

%-3.3-------------------------------------------------

\subsection{Conformal field theory of  Weyl fermions and 
ground-state conditions}

We now recall some basic facts of the free chiral fermionic 
theory on the geometry of the spacetime cylinder.
The Virasoro and current modes are 
written in terms of fermionic Fock space operators as follows \cite{CDTZ},
\ba
\label{viras}
\hat{L}_n&=&\sum_{k=-\infty}^{\infty}\left(k-\frac{n}{2}- \mu\right)
: \hat{d}^{\dagger}_{k-n}\hat{d}_{k}:,
\\
\label{modes}
\hat{\r}_n&=&\sum_{k=-\infty}^{\infty}
: \hat{d}^{\dagger}_{k-n}\hat{d}_{k}:,
\ea
where the normal ordering $:():$ subtracts the infinite contribution
of the Dirac sea. Let us conventionally fill the sea up to 
the $k=0$ state \cite{CDTZ}, i.e. define the conformal theory 
ground-state by the conditions:
\ba
&&\hat d_k\vert \W,\mu \rangle=0,\qquad k>0,
\nl
&&\hat d_k^\dag\vert \W,\mu \rangle=0,\qquad k \le 0.
\label{norm-ord}
\ea
Thus, the normal ordering is defined by putting $\hat d_k$ to the 
right of $\hat d^\dag_k$ for $k>0$ and viceversa for $k\le 0$.
The commutation relations of Virasoro and current modes 
(\ref{viras}, \ref{modes})
are then obtained from those of the Fock space, with the result \cite{CDTZ}:
\begin{equation}
\label{chi-algebra}
\begin{aligned}
&\left[\hat{\rho}_{n}, \hat{\rho}_{m}\right]=n\delta_{n+m,0},\\
&\left[\hat{L}_{n}, \hat{\rho}_{m}\right]=-m\hat{\rho}_{n+m},\\
&\left[\hat{L}_{n}, \hat{L}_{m}\right]=\left(n-m\right)\hat{L}_{n+m}+
\frac{1}{12}\left(n^{3}-n\right)\delta_{n+m,0}.
\end{aligned}
\end{equation} 
This is the well-known current algebra with central charge $c=1$ \cite{cft}.

The Hamiltonian of the conformal theory on the cylinder
is expressed in terms of the Virasoro generator $\hat L_0$,
\begin{equation}
\label{confo-ham}
\hat{H}=\frac{v}{R}\left(\hat{L}_{0}-\frac{c}{24}\right),
\qquad c=1,
\end{equation} 
and it includes the Casimir energy. Furthermore, the conformal dimensions
$h$, eigenvalues
of $\hat{L}_0$, determine the fractional statistics $2h$ and 
conformal spin $h$ of excitations through the two point functions.

As shown in Ref.\cite{CDTZ}, the chemical potential $\mu$ 
in the expression (\ref{viras})
plays a double role: it determines the boundary conditions
of the Weyl fermion on the edge, i.e. 
$\psi(\th=2\pi)=\exp(i2\pi\mu)\psi(0)$, and parameterizes 
the following ground-state expectation values:
\begin{align}
\label{cgs}
&\hat{L}_{n}\ket{\Omega,\mu}=\hat{\rho}_{n}\ket{\Omega,\mu}=0, 
\qquad n>0,\\
\label{gr-energy}
&\hat{L}_{0}\ket{\Omega,\mu} =\frac{1}{2}
\left(\frac{1}{2}- \mu\right)^2\ket{\Omega,\mu},\\ 
\label{gr-charge}
&\hat{\rho}_{0}\ket{\Omega,\mu} =\left(\frac{1}{2}-\mu\right)\ket{\Omega,\mu}.
\end{align}
These values of charge and Virasoro dimension amount to 
finite renormalization constants  
that should be added to the definitions of $\hat {L}_0$ and $\hat{\r}_0$ in
(\ref{viras}, \ref{modes}). The two values are related among themselves 
by the fulfillment
of the current algebra (\ref{chi-algebra}), and can actually 
be computed by checking the commutation relations on the expectation values 
$\bra{\W,\mu}\hat{L}_n\hat{L}_{-n}\ket{\W,\mu}$ and 
$\bra{\W,\mu}\hat{L}_n\hat{\r}_{-n}\ket{\W,\mu}$ \cite{cft}. 

Conformal invariance of the ground-state requires that the values in
Eqs.(\ref{gr-energy}, \ref{gr-charge}) vanish and thus the chemical
potential is dynamically tuned to $\mu=1/2$, corresponding to standard
antiperiodic boundary conditions for fermions.  Other values of $\mu$
are possible, but they have specific physical meaning: for example,
$\mu=0$ for periodic fermions corresponds to another (Ramond) sectors
of the theory, that is non-perturbatively related to the
(Neveu-Schwarz) antiperiodic sector.
In the following, the values $\mu=1/2+\Z$ will also be considered
for realizing features of the dynamics of edge states.

%-4---------------------------------------------
\section{Orbital spin in the edge theory}\label{4}

In section 3, we saw that the effective theory predicts
non-vanishing values for charge and spin at the edge of the disc,
respectively proportional to the average orbital spin $\ov{s}$ and
average square $\ov{s^2}$, Eqs.(\ref{t-spin}), (\ref{t-charge}).
In this section, we are going to discuss how these effects
can be realized  in the microscopic description of
the edge for integer fillings in section 3.
These ground-state values of charge and spin follow from a kind of 
Casimir effect in the relativistic conformal theory, after careful discussion
of the (re)normalization conditions.
As a matter of fact, we show that these effects are rather non-generic
in the integer Hall effect with dynamic edge excitations.
On the other hand, in the case of fractional fillings discussed in section 5,
we shall argue that such ground-state values should be expected.

%-4.1---------------------------------
\subsection{Qualitative boundary conditions}

In compact spatial geometries, the linear relation between Landau
level degeneracy and flux is corrected by a finite amount, the
so-called shift, proportional to the orbital spin $s_n$, as shown by
Eq. (\ref{deg-flux}).  For the integer Hall effect on
the sphere, the degeneracy is given by $D_n=N_\phi+ 2s_n =N_\phi+ 2n+1$, where
$n=0,1,\dots$ is the level.

\begin{figure}
\centering
\includegraphics[scale=0.6]{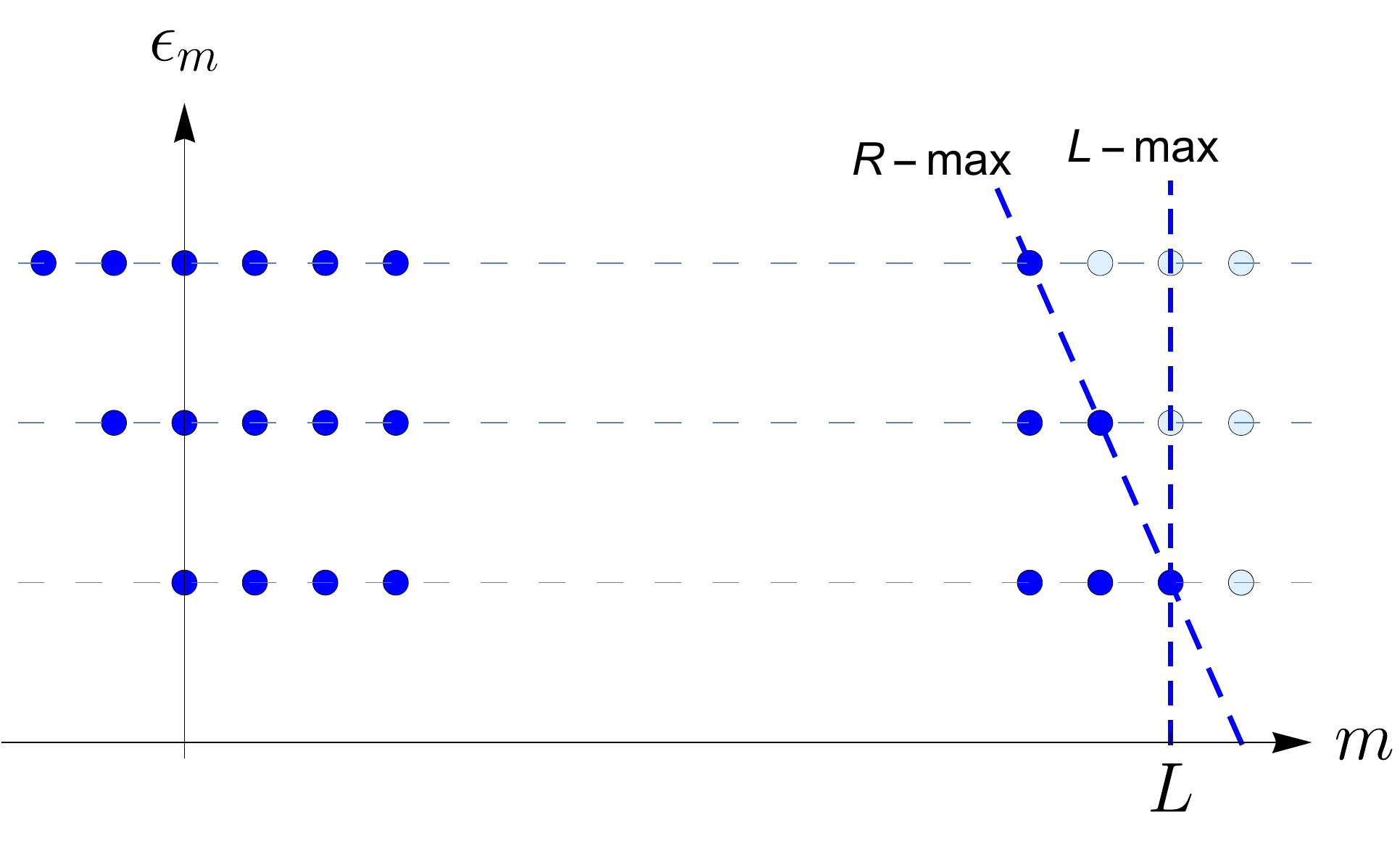} 
\caption{$L-max$ versus $R-max$ qualitative boundary conditions.}
\label{fig2}
\end{figure}

In the case of the disk,  half of this
correction (apart from the $1/2$) can indeed be realized when the levels
are filled up to a common value $L$ of angular momentum, leading to
$D_n=L+n$ (see Fig.\ref{fig2}).  This truncation of the spectrum,
dubbed $L-max$, can ideally be obtained by cutting the sphere in two
discs. The question we want to address in the following 
is whether this boundary condition can be realized
in a physical boundary with edge excitations.

The analysis in section 3 of wave functions with
momenta $m$ near $L$, $m=L+m'$, Eq. (\ref{nLL-edge}), 
shows that they are Gaussian peaked at positions $x=(m'+n)/2R$,
with $r=R+x$, $R^2=L$. This implies that for a common value of momenta,
e.g. $m'=0$, the states of higher levels are slightly displaced outward by 
$\Delta x=n/R$. In presence of boundary conditions due to
a confining potential or a maximum spatial extension, 
dubbed $R-max$, such states acquire higher energies. 
Therefore, we may expect that the filling of levels up to a common
Fermi energy could imply $m\le L-n$, leading to equal filling
$D_n=L$ for each level (see Fig.\ref{fig2}).  In this case, the shift
predicted by the effective action of section 2 would be washed out.

In conclusion, a `geometric' boundary condition of the kind $L-max$ would
realize the prediction of Eq. (\ref{deg-flux}),
while a `dynamic' condition of the kind $R-max$ would give no effect.
In the following, we describe the detailed realization of these two cases.

%-4.2--------------------------------------------

\subsection{Edge spectrum}

The chiral conformal field theory of the Weyl fermion possesses a
linear spectrum of edge excitations of the form \cite{CDTZ}:
\be
\eps(k)=v (k-k_F-\mu)=\frac{v}{R}(m'-\mu), \qquad |m'|<O(\sqrt{L}),
\quad L=R^2,
\label{lin-spectrum}
\ee
where $k$ is the edge momentum, $v$ and $k_F$ are the Fermi velocity 
and momentum, $\mu$ is the chemical potential and $m'$ 
the angular momentum with respect to the value $L$ near the boundary.
Indeed, by inserting this spectrum in the second-quantized fermion Hamiltonian,
one reproduces the expected form $\hat H = v \hat L_0/R + {\rm const.}$
(\ref{confo-ham}) with the Virasoro generator given by (\ref{viras}).

We now discuss the confining potentials that can lead to a linear spectrum. 
We consider `macroscopic' 
potentials of the form $V(r)\sim r^k/R^{k-1}$, with $k=1,2,\dots$,
that are smooth functions of the coordinate $r$ 
 and have a linear term in the expansion
near the edge $r=R+x$, $x=O(1)$ for $R\to\infty$.
Upon subtracting the infinite term, the expansion near the edge
takes the form:
\be
V(x,R)=a_1 x + a_2 \frac{x^2}{R} + a_3 \frac{x^3}{R^2}+ \cdots,
\label{spec-x}
\ee
where the coefficients $a_i$ are dimensionless numbers
(the common dimensional scale $1/\ell^2$ is implicit).
Other forms of the potential polynomial in $x$ would introduce
dimensional constants that break scale invariance:
they correspond to non-relativistic corrections
to the conformal theory that are disregarded here (see Appendix A).

Furthermore, the potential should grow monotonically, because
oscillations would create additional chiral-antichiral pairs of edge
modes that interact and become massive, as analyzed in the studies of
edge reconstruction \cite{edgereco}. These oscillations, expanded in
the basis of Hermite polynomials $V =\sum_n b_n H_n(x)$ could single
out one edge from the others, as discussed in section 3.

The determination of the energy spectrum of the Landau levels with
potential (\ref{spec-x}) can be done analytically in the limit
$R\to\infty$ and the result is the following (see Appendix B). The
terms $O(x^k/R^{k-1})$ with $k=3,4,\dots$ in (\ref{spec-x}) yield
subleading corrections with respect to $O(1/R)$, and one is left with
a one-parameter family of potentials corresponding to the linear and
quadratic terms. The corresponding spectrum is:
\be
\eps_{n,m'}=2n+1+\frac{v}{R}\left(m'+ n (2+b) \right) +{\rm const.},
\qquad |m'|<O(R),
\label{spec-m}
\ee
where $n=0,1,\dots$ is the Landau level index, $(2n+1)$ is the bulk energy
and the constants are $a_1=2v$ and $a_2=v (1+b)$. In particular,
$b=0$ corresponds to the simpler quadratic potential $V=vr^2/R$.

The result (\ref{spec-m}) shows that in higher Landau levels 
the spectrum at fixed momentum $m'$ is shifted upward by an amount $O(n/R)$ 
as anticipated by the qualitative argument at the beginning of this chapter.
This is the effect of the orbital spin at the edge, more precisely
of the differences $s_n-s_{n-1}=n$ between levels, because a constant
term can be added at will.
This result is not completely universal, i.e. independent of the form 
of boundary potentials compatible with scale invariance at the edge,
since it involves one free parameter $b$, corresponding to the
quadratic correction to linearization.

In conclusion, the microscopic Hamiltonian of the $\nu=n$ Hall effect 
with boundary potential takes the form:
\be
\begin{aligned}
\hat {H} &= \hat {H}_0 +\frac{v}{R}\int dz^2
\hat{\Psi}^\dag\, V(x,R)\, \hat{\Psi}+ \text{const. }
\\
&=\sum_{i=0}^n \sum_{m_i\in\Z}\left[ 2i+  \frac{v}{R}
\left(m_i+s_i(2+b) -\a\right)\right]
\hat{d}^{(i)\dag}_{m_i}\hat{d}^{(i)}_{m_i}
+{\rm const.}\;.
\label{op-ham}
\end{aligned}
\ee
In this equation, $\hat {H}_0 $ is the bulk Hamiltonian
(\ref{landau-ham}), the momenta $m_i$ are measured w.r.t. $L=R^2$ and
the constant term $\a$ is put explicitly.

The correction to the linear edge dispersion relations (\ref{op-ham})
due to the shifts $s_i$ is rather relevant for the following
discussion. We remark that this effect cannot be modified, apart from the
overall constant $\a$. Moreover, in the geometry of the annulus, the
antichiral edge modes of the inner circle possess the same spectrum
(\ref{op-ham}) with $v\to |v|$ and same shifts.

%-4.3------------------------------

\subsection{Map to conformal field theory}

We now identify the edge Hamiltonian (\ref{op-ham}) with
the conformal theory form (\ref{viras}), 
i.e. compare the following two expressions level by level, 
\ba
\label{op-ham-i}
\hat{H}^{(i)}&=& \frac{v}{R}
\sum_{m_i\in\Z}\left[2i+
 \frac{v}{R}\left(m_i+\frac{(2i+1)(2+ b)}{2}-\a\right) \right]
\hat{d}^{(i)\dag}_{m_i}\hat{d}^{(i)}_{m_i},
\\
\frac{v}{R}\hat{L}_0^{(i)}
&=&\frac{v}{R}\sum_{k_i\in\Z}\left(k_i-\mu_i\right)
:\hat{d}^{(i)\dag}_{k_i}\hat{d}^{(i)}_{k_i}:\;, \qquad i=0,1,\dots.
\label{viras-i}
\ea
Note that the conformal Hamiltonian does not include the bulk energy,
that is assumed to be constant for edge physics, because
bulk-boundary electron transitions are disregarded.
The conformal description is robust to deformations that set independent
Fermi velocities for each level, $v\to v_i$, being non-universal
parameters; in the following, we keep a single velocity without loss of
generality.

This matching of the two Hamiltonians (\ref{op-ham-i}),(\ref{viras-i})
involves two aspects:
\begin{itemize}
\item
The $2s_i=2i+1$ shift in the dispersion relation 
can be accounted for by assuming level-dependent values for
the chemical potential $\mu_i$ in the conformal theory description
of the $i$-th level. This setting leads to physically 
observable effects of the orbital spin.
\item
The conformal mode index $k_i$ in (\ref{viras-i}) can also be 
translated w.r.t. the edge momentum $m_i$, in order to match the
actual filling of the $i$-th Landau level with the conformal vacuum
(\ref{norm-ord}), conventionally filled up to level $k_i=0$.
\end{itemize}

Let us first identify the conformal theory for the
lowest Landau level, $i=0$.
For $N=L+1$ electrons, the ground-state is filled up to level $m=L$, 
thus the conformal moding and edge momentum precisely match, $k_0=m_0$.
The conformal invariant value $\mu=1/2$ for the chemical potential
is fixed by adjusting the constant $\a$ of the confining potential \ref{op-ham},
as follows:
\be
\mu_0=\frac{1}{2}, \qquad \a=\frac{3}{2}, \qquad \left(k_0=m_0 \right).
\ee

The determination of the conformal theories for higher levels
depends on the physical setting. We can distinguish two
cases, that are analyzed in the following paragraphs.

%-4.3.1-------------------------------------------------

\subsubsection{Smooth boundary}

\begin{figure}
\centering
\includegraphics[scale=0.4]{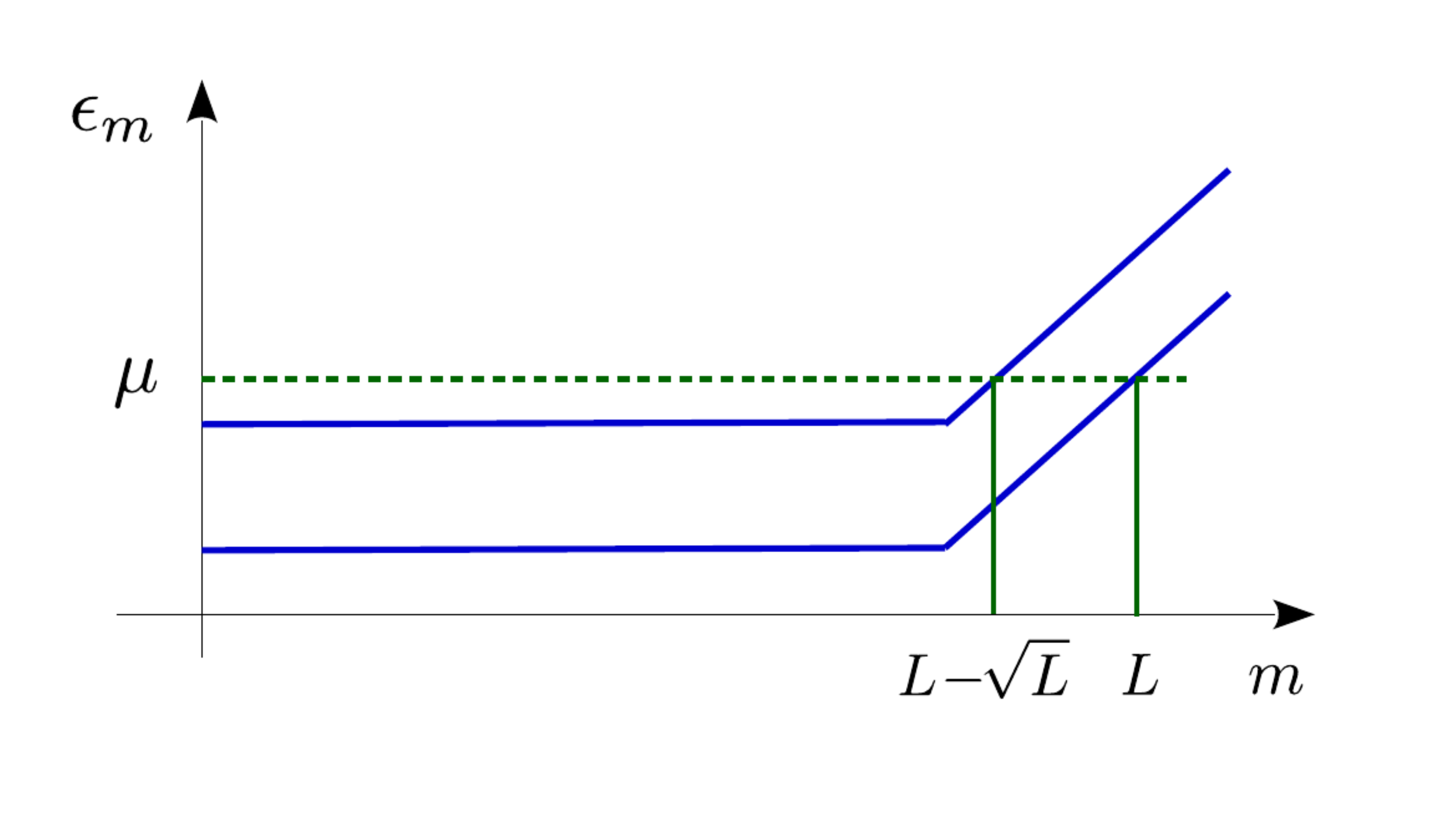} 
\caption{Filling of two Landau level according to the spectrum 
(\ref{spec-m}).}
\label{fig3}
\end{figure}

Let us consider the filling of the first two Landau levels 
up to a common Fermi energy, in presence of the boundary potential
(\ref{spec-x}) (see Fig.\ref{fig3}). The comparison of energies 
(\ref{spec-m}), $\eps_{0,m_0}=\eps_{1,m_1}$, gives:
\be
m_0=\frac{2R}{v} + m_1+2 +b .
\label{fermi-match}
\ee
Being $m_0=0$ for the lowest-level ground-state, this equation shows
that the second edge branch is located around momenta $m_1\sim - 2R/v$.
Therefore, the corresponding conformal theory
should be defined with respect to shifted momenta, as follows:
\be
k_1=m_1-2R/v-2-b, \qquad \mu_1=\frac{1}{2}.
\ee
Recalling that $R=\sqrt{L}$ is large, this can be adjusted by $O(1)$
corrections such that the difference $k_1-m_1$ is an integer and 
independent of $s_1$, in particular.
The chemical potential $\mu_1$ is again fixed by requiring
vanishing ground-state charge and conformal spin.

This is the realization of the $R-max$ boundary condition
described at the beginning of this section: the effect of the shift
is cancelled because the conformal field theories of the two edge branches are
defined independently one of the other. Note that the relative
wavefunctions are at distance $\Delta m \sim 2R$, i.e. 
$\Delta x\sim 2$ and thus have exponentially small overlaps.
There are no particle exchanges between the two edges in agreement with
the independent conservation of the relative charges.

In conclusion, for isolated droplets with smooth confining potentials
the edge excitations of different levels are orthogonal and the system
realizes the $R-max$ qualitative boundary setting, leading to
no orbital spin effects.
We have verified that the same behavior is present in the
Landau level spectrum with Dirichlet boundary conditions at $r=R$.
Finally, when the system is connected to an electron reservoir, the edge
branches are let to interact, but their chemical potentials level off and
there is no effect either.

%-4.3.2-------------------------------------------------

\subsubsection{Sharp boundary}

We now discuss the realization of the $L-max$ sharp boundary condition
introduced at the beginning of this section. 
We must assume that the bulk energy of Landau levels in
(\ref{spec-m}) is absent; this could be due to the bulk interactions
of electrons in the incompressible fluid, that were not included
in the previous microscopic theory. For example, there could be a
level-dependent screening of the Coulomb interaction,
but we have no proof of this fact. Let us nonetheless analyze the
map to conformal field theory in this setting.

The dispersion relations $\eps_{n.m}$ for vanishing bulk term
show that the different branches of edges occur in the same $O(1/R)$
range of energies and for close momenta $m_i$ values, such that
the corresponding wavefunctions do overlap.
The $n$ branches are now parts of the same $c=n$ conformal theory and it
necessary to define a unique ground-state with a common definition of
edge momentum. This is given by:
\be
k_i=m_i=m_0, \qquad\quad i=1,2,\dots,
\label{m-sharp}
\ee
such that all levels are filled up to $m_i=0$.
This ground-state is translation invariant for all the levels;
otherwise said, there are no persistent currents at the edge.

\begin{figure}
\subfigure[\label{fig4a}]{
\includegraphics[scale=0.65]{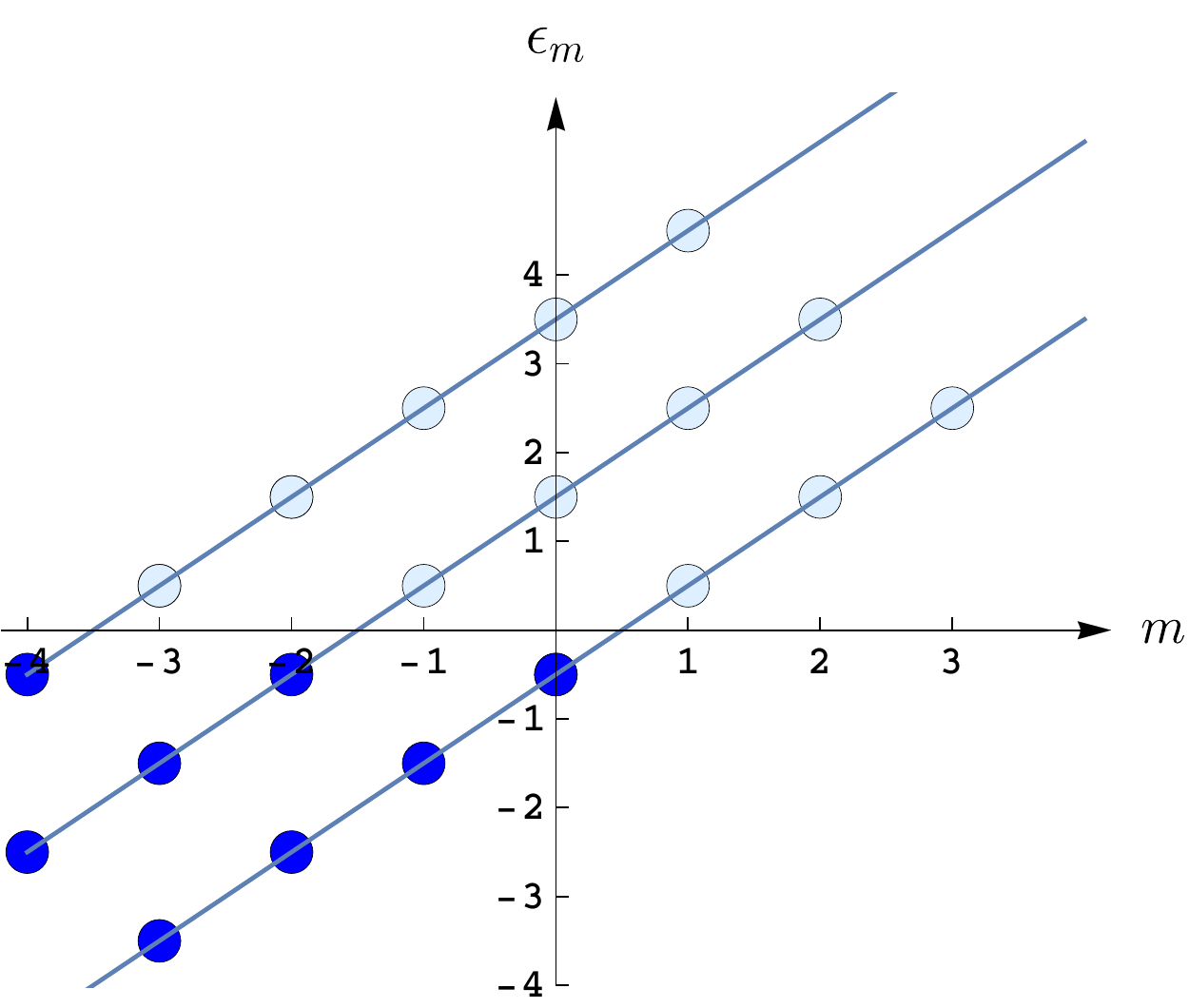} 
}
\subfigure[\label{fig4b}]{
\includegraphics[scale=0.65]{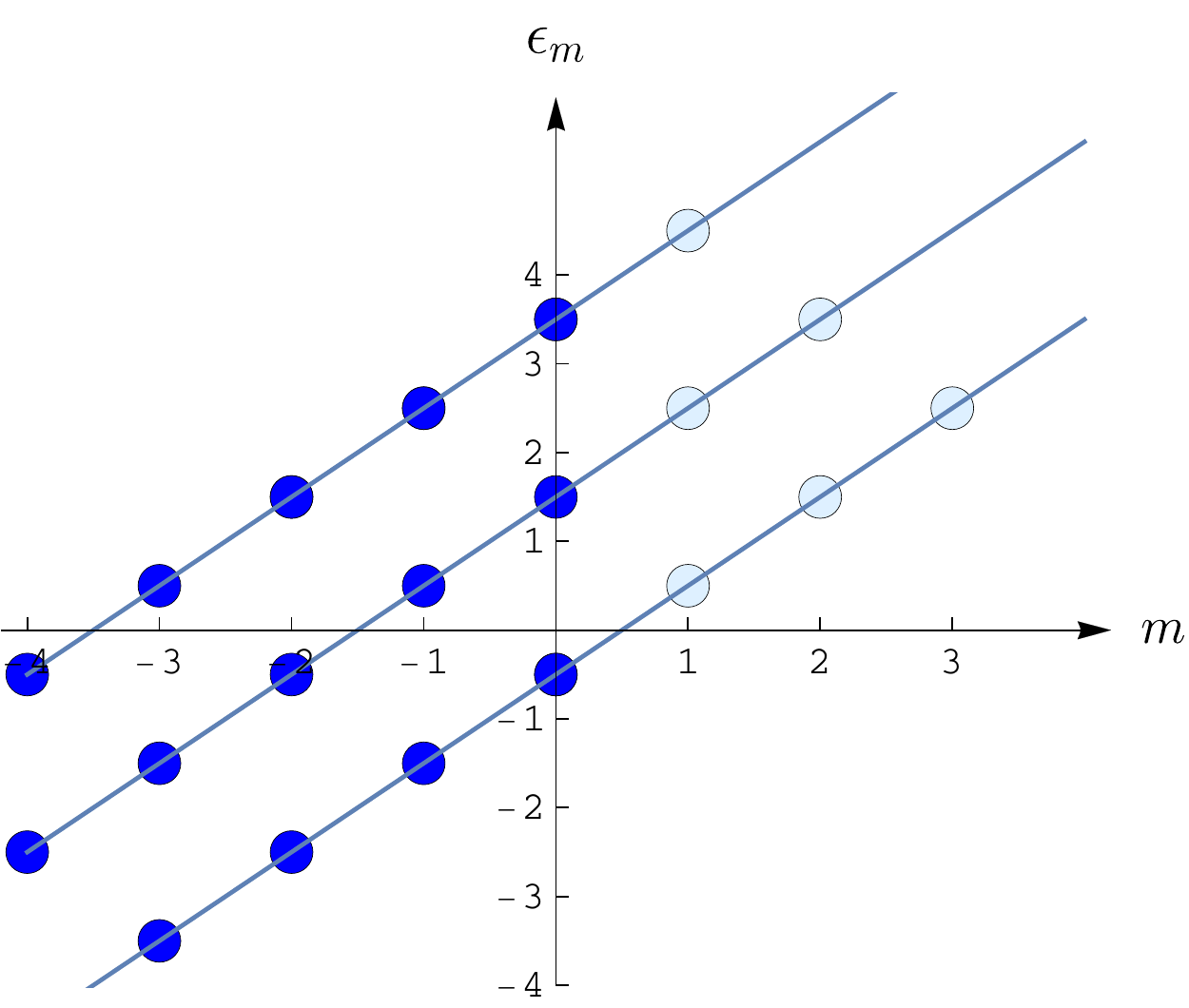} 
}
\caption{Fock space near the Fermi surface of the $\nu=3$ droplet in
  two different setups: (a) all levels with same chemical potential,
  $\mu_{i}=1/2$, corresponding to vanishing ground-state energy;
  (b) all levels filled up to the same momentum $m_{i}=0$. }
\end{figure}

The matching of microscopic and conformal Hamiltonians (\ref{op-ham-i}), 
(\ref{viras-i}) is achieved in this case by allowing a different chemical 
potential for each level (we set $b=0$ momentarily):
\be
\mu_i=\frac{1}{2}-2i, \qquad i=1,\dots,n-1.
\ee
These values of $\mu_i$ imply that 
the higher Landau levels possess non-vanishing ground-state values 
of charge and conformal spin (dimension), as described in section 3,
Eqs. (\ref{gr-energy}, \ref{gr-charge}):
\begin{align}
\label{gr-charge-1}
&\hat{\rho}_0^{(i)}\ket{\Omega,\mu_i} = 2i\ \ket{\Omega,\mu_i},
\qquad\quad \mu_i=\frac{1}{2}-2i,
\\
&\hat{L}_{0}^{(i)}\ket{\Omega,\mu_i} =2 i^2\ 
\ket{\Omega,\mu_i}.
\label{gr-energy-1}
\end{align}
It turns out that these states are actually excitations with respect of the 
standard vacuum with $\mu=1/2$. 
This realization of the $L-max$ boundary condition
is shown in Fig.\ref{fig4b} next to the earlier
$R-max$ condition in Fig.\ref{fig4a}.

The total ground-state charge and conformal spin  for 
$\nu=n$ are given by the sum of the contributions of all the levels:
\ba
Q_b&=&\sum_{i=0}^{n-1} 2i=n(n-1),
\nl
{\cal S}_b &=&\sum_{i=0}^{n-1}2i^2= \frac{n(n-1)(2n-1)}{3}.
\label{gr-edge}
\ea
These also imply the ground-state energy:
\be
E_b=\frac{v}{R}\left({\cal S}_b  - \frac{c}{24}  \right), \qquad
c=n.
\ee
This is the anticipated Casimir effect due to the orbital spin.

Let us add some remarks:
\begin{itemize}
\item
The ground-state values (\ref{gr-charge-1}), (\ref{gr-energy-1})
 acquire factors, respectively, 
$(1+b/2)$ and $(1+b/2)^2$ for general quadratic terms in the confining 
potential (\ref{spec-x}). Thus the Casimir effect is not fully universal.
\item
However, the allowed values of the chemical potential 
are $\mu_i=1/2+\Z$, corresponding to 
antiperiodic boundary conditions for fermions; other real values of $\mu_i$
would cause unphysical non-analyticities in electron correlators.  In
other words, the charge accumulated in the ground-state $Q_b$
should be an integer. This implies that only $b\in\Z$ is
compatible with conformal invariance at the edge.
A mechanism for self-tuning of $b$ is the quadrupole deformation of the
droplet discussed in section 6.
\item
The ground-state values (\ref{gr-charge-1}), (\ref{gr-energy-1}) agree
with the effective field theory results (\ref{t-charge}),
(\ref{t-spin}) for the special case $b=-1$.
\end{itemize}

In conclusion, we have found that the edge effects parameterized by
the orbital spin can indeed been found in the microscopic description
when a unique conformal theory  encompasses all edges
branches. This definition of edge theory is non-generic in the case of
the integer Hall effect, where separate branches are possible, but it
is necessary for describing interacting edges, such as those of the
hierarchical Jain states discussed in the next section.

%-5--------------------------------------------- 
\section{$W_\infty$ symmetry and fractional fillings}\label{5}

In this section, the edge theory with integer fillings is
rederived by using the symmetry of incompressible Hall fluids under
area-preserving diffeomorphisms of the plane, the $W_{\infty}$
symmetry \cite{CTZ1} \cite{Sakita} \cite{CTZ2}. 
This reformulation allows us to extend our analysis to
fractional fillings and in particular to hierarchical states, which
also possess several branches of edge modes and corresponding $s_i$ values.

%-5.1-------------------------------------------------
\subsection{Edge excitations as $W_{\infty}$ transformations}

We consider the filled lowest Landau level as a starting point of 
our discussion. As is well known, the electrons form a droplet of 
incompressible fluid that is characterized by constant density and
constant area ${\cal A}$.
The deformations of the fluid at energies below the bulk gap
can be generated by coordinate transformations
of the plane that keep the density constant in the bulk. 
These corresponds to the area-preserving 
diffeomorphisms that modify the shape of the droplet at fixed area 
\cite{CTZ1}\cite{Sakita}\cite{CTZ2}.

At the classical level, these reparameterizations are expressed in terms
of a  scalar generating function $w(z,\bar z)$ and 
Poisson brackets. Their action on the coordinate and the
density $\r$ is given by the following expressions:
\be
\d_\w z=\{z,w\}=\eps^{z\bar z}\de_z z\de_{\bar z} w + 
(z \leftrightarrow \bar z), \qquad
\d_w\r=\{\r,w\}, 
\label{c-fluct}
\ee
where $\eps^{z\bar z}= -\eps^{\bar z z}=-2i$.
In the geometry of the disk, the ground-state density is constant in
the bulk and goes to zero at the boundary: the deformation by Poisson
brackets (\ref{c-fluct}) involves the derivative of the density that
has support on the boundary, as expected.

At the quantum level, the quantities in (\ref{c-fluct}) become one-body 
operators, 
\be
\hat\r=\hat\Psi^\dag\hat\Psi,\qquad
\hat w =\int d^2z \hat\Psi^\dag (z,\bar z)w(z,\bar z)\hat \Psi(z,\bar z),
\ee
involving the field operator $\hat\Psi$ restricted to 
the first Landau level. The Poisson brackets (\ref{c-fluct}) 
are replaced by commutators, $\d \hat\r= i\left[\hat \r, \hat w \right]$.
The ground-state expectation value gives the transformation of the 
density function, that takes the following form \cite{CTZ2} \cite{CR}:
\begin{equation}
\label{Moyal}
\delta \rho\left(z,\bar{z}\right)=
\bra{\W}\left[\hat\r(z,\bar z),\hat w\right] \ket{\W}=
i\sum_{n=1}^{\infty}
\frac{\left(2\hslash\right)^{n}}{B^{n}n!}
\left(\partial_{\bar{z}}^{n}\rho\partial_{z}^{n}w-
\partial_{z}^{n}\rho\partial_{\bar{z}}^{n}w\right)=
\left\lbrace\rho,w\right\rbrace_{M}.
\end{equation}
This formula defines the Moyal brackets $\{\r,w\}_M$, that are 
non-local due to the non-commutativity of coordinates in the lowest
Landau level. There appears an expansion in $\hbar/B$, whose first term
reproduces the classical transformation law (\ref{c-fluct}).

Equation (\ref{Moyal}) expresses the $W_\infty$ transformations of the density
at the quantum level.
Another formulation of this symmetry involves the Girvin-MacDonald-Platzman 
sin-algebra \cite{GMP}, that corresponds to
the commutator of two densities in Fourier space: similar to Eq.(\ref{Moyal}), 
this algebra is given by the Moyal brackets of the classical densities.

In the following, we discuss the form of the leading $O(1/B)$
term in the Moyal brackets, while
the higher orders $O(1/B^k)$ will be briefly discussed in appendix B.
The density $\rho^{(0)}$ of the filled lowest level in the disk geometry 
is a function of the radius only, $\rho^{(0)}=\rho^{(0)}\left(r\right)$.
Thus, the leading $W_\infty$ deformation of the ground-state can be written 
as follows:
\be
\delta\rho^{(0)}(r,\th)=
\frac{2i}{B}\;\bar{\partial}\rho^{(0)}\;\partial w+h.c.
=\frac{1}{rB}\left(\partial_{r}\rho^{(0)}(r)\right)
\partial_{\theta}w(r,\theta).
\label{1LL-Moy}
\ee

We now define the Fourier modes of the edge density by integrating in
space the bulk density, following the same steps of the analysis in
section 3,
\begin{equation}
\label{def-mom}
\delta\rho_{k}^{(0)}=\int d\theta\,e^{-ik\theta}\int drr\,\delta\rho^{(0)}.
\end{equation}
It is also convenient to expand the generating function 
in Fourier modes, leading to ($B=2$ hereafter):
\be
\delta\rho_{k}^{(0)}=\frac{ik}{2}\int dr \left(\de_r\r^{(0)}(r)\right)
w_k(r), \qquad
w\left(r,\theta\right)=\frac{1}{2\pi}\sum_{n\in\Z} 
w_{n}\left(r\right)e^{in\theta}.
\ee
The remaining integral over the radial dependence 
is non-vanishing in a shell $r=R\pm O(1)$, as understood in section 3. 
In order to compute the integral, 
we can use the exact expression for the derivative of the 
density of filled Landau levels derived in Ref.\cite{Dunne}.
For the $i$-th level filled with $N$ electrons, it reads:
\be
\frac{d}{dr^2}\r^{(i)}(r)=
- i! \;\frac{e^{-r^2}r^{2N-2i-2}}{(N-1)!}L_i^{N-i-1}(r^2)\; L_i^{N-i}(r^2).
\label{Dunne}
\ee
The expression for the lowest level is evaluated in the limit to the edge
defined in section 3, finding the result, 
\be
\de_r\r^{(0)}(R+x)\simeq -2 e^{-2x^2}, \qquad R\to\infty, \quad x=O(1).
\label{edge-lim-den}
\ee
Thus, it is again useful to use Hermite polynomials
$H_{2n}(2x)$ for analyzing the radial dependence
of edge excitations.
We obtain:
\be
\d\r^{(0)}_k=-ik\; w_{0k},\qquad 
w(r,\th)=\frac{1}{2\pi}\sum_{n\in\Z}\sum_{i=0}^\infty w_{ik}\; 
H_{2i}(\sqrt{2}x)\; e^{in\th}.
\label{1LL-fluct}
\ee
This classical amplitude of fluctuations should be compared with the simplest 
excitation of the $c=1$ conformal theory, that is
given by the current-algebra mode applied to the ground
state, $\ket{ex}\sim \hat\r^{CFT}_{-m}\ket{\W}$. 
Therefore, the conformal theory analog of the $W_\infty$ density fluctuation
(\ref{1LL-fluct}) reads,
\begin{equation}
\label{edge-fluct}
\delta\rho^{CFT}_{k}=i\langle\Omega\vert\left[\hat{\rho}^{CFT}_{k},
\hat{\rho}^{CFT}_{-m}\right]\vert\Omega\rangle= i\delta_{km}k.
\end{equation}
The equivalence of the results (\ref{1LL-fluct}) and (\ref{edge-fluct})
directly shows the relation between the bulk $W_{\infty}$ symmetry and
the edge conformal symmetry. 

The $W_{\infty}$ deformations in higher filled levels are described
in similar fashion. The transformations act horizontally within each
level \cite{CTZ1}, that can be analyzed independently; this matches the fact
already discussed that the corresponding branches of edge excitations
are orthogonal.
The fluctuations of the second level are given by:
\be
\label{2LL-fluct}
\delta\rho^{(1)}=i\langle\Omega^{(1)}\vert\left[
\hat{\rho}^{(1)},\hat{w}^{(1)}\right]\vert\Omega^{(1)}\rangle,
\end{equation}
where $\hat{\rho}^{(1)}=\hat\Psi^{(1)\dag} \hat\Psi^{(1)}$
is the corresponding density operator, and the generator is expressed as:
\begin{equation}
\label{w1}
\hat{w}^{(1)}=\int d^2z\,\hat{\Psi}^{(1)\dagger} 
w\left(z,\bar{z}\right)\hat{\Psi}^{(1)},
\end{equation}
in terms of the field operator $\hat{\Psi}^{(1)}$ restricted 
to the second level.
We now remark that expressions like (\ref{2LL-fluct}, \ref{w1}) 
can be mapped into lowest level formulas 
by using the relations among wavefunctions (\ref{Landau-wfunct}), such as 
$\psi_{1,m}=a^\dag \psi_{0,m+1}$.
This corresponds to a differential relation between the 
field operators of the two levels, leading to:
\begin{equation}
\label{rel-1-2LL}
\hat{\rho}^{(1)}= \left(1+\partial\bar{\partial}\right)\hat{\rho}^{(0)},
\end{equation} 
and
\begin{equation}
\hat{w}^{(1)}=\int d^{2}z\, \hat{\Psi}^{(0)\dagger}
\left[\left(1+\partial\bar{\partial}\right)w\right]\hat{\Psi}^{(0)},
\end{equation} 
implicitly assuming the isomorphism of the two Fock spaces.
The corresponding relations among classical functions are:
\begin{equation}
\rho^{(1)}=\left(1+\partial\bar{\partial}\right)\rho^{(0)}, 
\qquad w^{(1)}=\left(1+\partial\bar{\partial}\right)w.
\end{equation}
These differential relations map the second level density fluctuations
(\ref{2LL-fluct}) into expressions that are defined on the first level,
for which the Moyal brackets (\ref{Moyal}) apply. Moreover, the two
kinds of differential relations commute.
Therefore, the leading $O(1/B)$ second level fluctuations are 
of the same form (\ref{1LL-Moy}):
\begin{equation}
\delta\rho^{(1)}=\frac{1}{rB} \partial_{r}\rho^{(1)}\;
\partial_{\theta}w^{(1)}.
\label{2LL-Moy}
\end{equation}
The edge moments are again defined by Fourier analysis in $\th$ and
radial integration as in Eqs. (\ref{den-modes}, \ref{def-mom}).  For
the derivative of the density we can consider the $R\to\infty$ limit
of the exact formula (\ref{Dunne}) or use the differential relation
(\ref{rel-1-2LL}), leading to
$\de_r\r^{(1)}=\de_r(1+\de\bar\de)\r^{(0)}\simeq -8x^2 e^{-2x^2}$.
Using this result together with (\ref{2LL-Moy}) and (\ref{def-mom}),
we obtain:

\begin{equation}
\delta\rho^{(1)}_{k}\simeq -ik\int dx\, 8x^2 e^{-2x^2}
\left(1+\de\bar\de\right)\sum_{i=0}^\infty w_{ik}\; H_{2i}(\sqrt{2}x).
\label{2LL-fluct.2}
\end{equation}
This fluctuations involve contributions from the first three radial moments
$(w_{0k},w_{1k},w_{2k})$. In the same way as the conformal field
theory (operator) description of sections 3,
the edge excitations of higher levels are associated to higher 
radial moments of the bulk density evaluated in the region $r=R+x$,
with $x= O(1)$.

The analysis can be extended to the third Landau level: by using the
relation between wavefunctions, we can again establish a differential
map to the first level:
\begin{equation}
\hat\rho^{(2)}=\left[1+2\partial\bar{\partial}+
\frac{1}{2}\left(\partial\bar{\partial}\right)^{2}\right]\hat{\rho}^{(0)}.
\end{equation}
Together with the analogous one for the generator $\hat w^{(2)}$,
one can compute the $W_\infty$ deformations of the ground-state
density (\ref{Moyal}) whose radial moments involve further terms of 
the radial expansion in Hermite polynomials.

The combination of the previous results finally
yield the edge excitations for integer filling fraction obtained from
the $W_\infty$ transformations of incompressible ground-states.
For example, in the $\nu=2$ case, we find using (\ref{1LL-fluct}) 
and (\ref{2LL-fluct.2}):
\be
\d\r_k^{(\nu=2)}=\d\r_k^{(0)}+\d\r_k^{(1)}=
-ik\left(a_0w_{0k}+a_1w_{1k}+a_2w_{2k}\right),
\ee
where $(a_0,a_1,a_2)$ are numerical coefficients that parameterize the
radial dependence at the edge.

In conclusion, in this section we have shown that the $W_\infty$
transformations of the ground-state for integer fillings $\nu=n$
generate edge excitations that match the conformal field theory
description of section 3: in particular, the $n$ independent branches
of excitations are associated to different radial moments of the
density in a finite shell at the boundary, $r-R= O(1)$.  As discussed
in section 3, Eq.(\ref{singleout}), it is possible to single out one
particular component by integrating the density with specific radial
weights and using the orthogonality of Hermite polynomials.

%-5.2-----------------------------------------------
\subsection{Edge excitations and orbital spin for fractional fillings}

The $W_\infty$ description of edge excitations is particularly useful
because it holds for any incompressible fluid ground-state, including
fractional fillings.  Namely, the Moyal brackets (\ref{Moyal})
correctly give the low-energy excitations that can be divided in
several branches by studying the radial moments of the density.  In
summary, this approach provides the general and universal kinematics
of edge excitations.  In the integer filling case, it shows that the
description of edge states remains valid when bulk interactions are
introduced that preserve incompressibility, i.e. do not close the gap.
Furthermore, this approach extends directly to the fractional Hall
effect and provides a unique derivation of the edge conformal field
theory.

The difference between the integer and  fractional filling lies
in the energetics of excitations, i.e. in the form of the edge
Hamiltonian. In the integer case, a direct microscopic derivation was possible
as shown in section 4. For fractional states, the edge
dynamics is due to the many-body interactions and cannot be derived
analytically. The standard approach, based on
the effective action of section 2 and bosonization of the edge fermions
predicts that the Hamiltonian takes the Luttinger current-current 
form \cite{cft}, 
\be
\hat{H}=\sum_i\frac{v^{(i)}}{R}\left(\hat{L}_0^{(i)} - \frac{1}{24}\right),
\qquad
\hat L_0^{(i)}=\sum_{k\in\Z} \; :\hat\r^{(i)}_{-k} \hat\r^{(i)}_k:\; .
\label{bos-ham}
\ee
This conformal theory still possesses integer central charge
and the spectra of charges and conformal spins are given by the 
weight lattice with Gram matrix $K^{-1}_{(i)(j)}$ in Eq.(\ref{nusbars})
as follows:
\be
Q_b=\l^T K^{-1} t, \qquad
{\cal S}_b= \frac{1}{2}\l^T K^{-1} \l, \qquad\quad \l=(\l_1,\dots,\l_n),
\quad \l_i\in \Z,
\label{K-spec}
\ee
where $t=(1,1,\dots,1)$ is the so-called charge vector and $\l_i$
characterize the excitations.

Although this theory has not been completely proven, analytic and
numerical results confirm its predictions, such as that the Laughlin
states with $\nu=1/(q+1)$, with $q$ even, possesses a single branch,
while the hierarchical Jain states with $\nu=n/(qn+1)$ show $n$
branches.
The extensive phenomenology based on the composite fermion 
correspondence \cite{jain} between integer and hierarchical states let us
argue that the structure of branches found for integer fillings also
describe the Jain states.

Regarding the role of the orbital spins $s_i$, we cannot prove that
the analysis of section 4 extends to Jain states, having
no direct derivation of their effect on the energy spectrum.
Nonetheless, based on some reasonable assumptions, we present a 
self-consistent argument for the existence of ground-state 
values of charge and spin in the edge theory as
predicted by the effective theory of section 2.

Since the hierarchical states are interacting, their conformal
theory should be built around a single ground-state, imposing the
$L-max$ boundary conditions of Fig.\ref{fig4b}.
As discussed in section 4.3.2, the resulting ground-state 
is actually an excited state from the 
conformal theory point of view, corresponding to a number of 
electrons added to the standard conformal vacuum $\ket{\Omega}_{CFT}$,
obeying:
\be
\r_0\ket{\Omega}_{CFT}= L_0\ket{\Omega}_{CFT}=0 .
\label{cft-v}
\ee
Such ground-state
$\ket{\W,k,h_k}$ with charge $Q_b=k\in\Z$ and conformal spin 
${\cal S}_b=h_k$
has  the form:
\ba
\ket{\W,k,h_k}&=&\lim_{\t_k\to -\infty, \t_1>\t_2>\cdots>\t_k}
:V_e(\eta_k)\cdots V_e(\eta_1):\ket{\W}_{CFT}
\nl
&=& \lim_{\eta\to 0}\; V_{ke}(\eta)\ket{\W}_{CFT}.
\ea
In this expression, $\eta_j=\exp((\t_j+iR\th_j)/R)$, where $\t_j+iR\th_j$
are the coordinates of the edge spacetime cylinder and $V_e(\eta)$ is
the vertex operator for the electron field on the edge; the
normal-ordering $(:\ :)$ should be evaluated by fusing $k$ electrons
in the conformal theory, leading to the $Q_b=k$ field $V_{ke}$. Since
electrons have integer mutual statistics w.r.t. all the excitations of
the theory, the insertion of these fields at infinity does not cause
additional nonlocalities in edge correlation functions, that are
actually independent of the insertion points $\eta_j$.

Let us assume that this ground-state is realized and run the consistency check.
We first determine the orbital spins $s_i$ for the hierarchical 
states. In the multicomponent Wen-Zee action of section 2,
the parameters $\nu$, $\nu \ov{s}$ and $\nu \ov{s^2}$ are
given by the general expressions (\ref{nusbars}):
\be
\nu=t^T K^{-1} t, \qquad
\nu \ov{s} =t^T K^{-1} s, \qquad
\nu \ov{s^2} =s^T K^{-1} s,
\label{wz-shift}
\ee
upon inserting the $s=(s_0,s_1,\dots,s_{n-1})$ vector 
and the matrix $K=I+qC$, where $C$ is the 
$n\times n$ matrix  with all entries equal to one.
One finds:
\be
\nu=\frac{n}{nq+1}, \qquad
\nu \ov{s}=\frac{1}{nq+1}\sum_i s_i= \frac{n}{nq+1}\frac{q+n}{2},
\label{hiera-s}
\ee
with $n=1,2,\dots$ and $q=0,2,4,\dots$.
In the second equation, we also wrote the value $\ov{s}=(q+n)/2$
obtained from the angular momentum of the Jain wavefunctions \cite{jain} 
and Eq.(\ref{tot-moment}). The relation (\ref{hiera-s}) identifies 
the following values:
\be
s_i=\left(\frac{q+1}{2},\frac{q+3}{2},\dots,\frac{q+2n-1}{2}\right).
\label{hiera-si}
\ee
Namely, they take half-integer values and their differences
are integer, as expected.

The comparison of these bulk data with the edge conformal theory
involves two steps: 
i) The ground-state charge $Q_b$ and spin ${\cal S}_b$ should
correspond to electron excitations of the edge theory; ii) They should
be parameterized by integers $\l_i$ that are equal to the orbital
spin values (\ref{hiera-si}) up to a constant, $\l_i=s_i+\D$.

Let us verify that these two conditions can be met. The
electron excitations have the spectrum:
\be
Q_b=\L^T t, \qquad
{\cal S}_b= \frac{1}{2}\L^T K \L, \qquad\quad \L=(\L_1,\dots,\L_n),\quad
\L_i\in \Z,
\label{K-electr}
\ee
corresponding to the integer-charge subset of the general excitation
spectrum (\ref{K-spec}), i.e. $\l=K\L$.

Note the similarity between the expressions (\ref{K-spec}) for (fractional)
charged excitations  and those involving the $s_i$ (\ref{wz-shift})
that we want to reproduce.
Since the orbital spin values differ by integers, $s_{i+1}-s_i=1$,
we seek for solutions $\l_i$ with the same property $\l_{i+1}-\l_i=1$. 
Upon inspection, we find that they do exist and are given by:
\be
\l_i =\L_i+ q \sum_j\L_j,   \qquad \qquad \L_i=(0,1,2,\dots, n-1), 
\label{K-solu}
\ee
correctly obeying $\l_i=s_i+\D$.
Actually, these electron excitations are possible due to the particular form
of the matrix $K$ of hierarchical states.

The ground-state values of charge and spin are finally given by:
\ba
Q_b&=&\sum_i\L_i=\frac{n(n-1)}{2}, \qquad\qquad\quad \nu=\frac{n}{nq+1},
\nl
{\cal S}_b&=&\frac{n(n-1)(2n-1)}{12} +\frac{q}{2}
\left(\frac{n(n-1)}{2})\right)^2.
\label{K-gs}
\ea 
Note that the value of the integer charge is the same as in the
$\nu=n$ case (\ref{gr-edge}) (for parameter $b=-1$).  

In conclusion, we have found the expressions of the orbital spins
$s_i$ of hierarchical states (\ref{hiera-si}) and shown that their values
shifted by a common constant correctly parameterize ground-state
charge and conformal spin in the edge theory.

%-6---------------------------------------
\section{Signatures of the orbital spin at the edge}

A thorough discussion of the experimental setups that may led
to the observation of the orbital spin $s_i$ in edge physics
is beyond the scope of this work. In the following, we sketch two possible
settings where an effect could be seen, at least in principle.
We discuss the isolated Hall droplet (no conduction) 
with  sharp boundary conditions, discussed in section 4.3.2,
that can be realized with small droplets.

%-6.1---------------------------------------
\subsection{Coulomb blockade}

In the experiment of Coulomb blockade,
an electron tunnels into an isolated droplet at zero bias.
As discussed in Ref. \cite{Coulomb}, the energy of charged edge states can be 
continuously deformed by
changing the capacity of the droplet, i.e. by squeezing its area.
The energy of the $k$-electron excitation over the ground-state, 
as given by the eigenvalue of the Virasoro operator $L_0$, is
modified by the charging energy as follows:
\be
E_k=\frac{v}{R}\frac{\left(k-\s \right)^2}{2}, \qquad \nu=1,
\ee
where $\s$ is the change of flux quanta due to the squeezing of the area.
When $\s=1/2$, the energies of the ground and one-electron states
become degenerate, $E_0=E_1$, and an electron can tunnel inside the droplet at
zero bias, causing a peak in the conductance, i.e. $\D Q=1$. 
Successive peaks
are found when $\s$ passes the values $1/2+j$, for $j=1,2,\dots$.

In the case of the isolated droplet with sharp boundary 
discussed in section 4.3.2, the electrons of the $i$-th level
possess higher activation energies for larger $i$ 
values, owing to the different chemical potentials. 
The previous equation is modified into:
\be
E_k^{(i)}=
\frac{v}{R}\left[\frac{\left(k-\s \right)^2}{2} +2i(k-\s)\right]=
\frac{v}{2R}\left[\left(k-\s+2i \right)^2 -4i^2\right], 
\qquad i=0,\dots,n-1.
\ee
This equation shows that for the $i$-th level, the first degeneracy point
$E_0^{(i)}=E_1^{(i)}$ occurs at the value $\s=1/2+2i$, while
the following ones repeat at the same distance $\D \s=1$.
Namely, the different branches of edge states enter into play 
at different $\s$ values, owing to the different activation energies.

\begin{figure}
\centering
\includegraphics[scale=1.2]{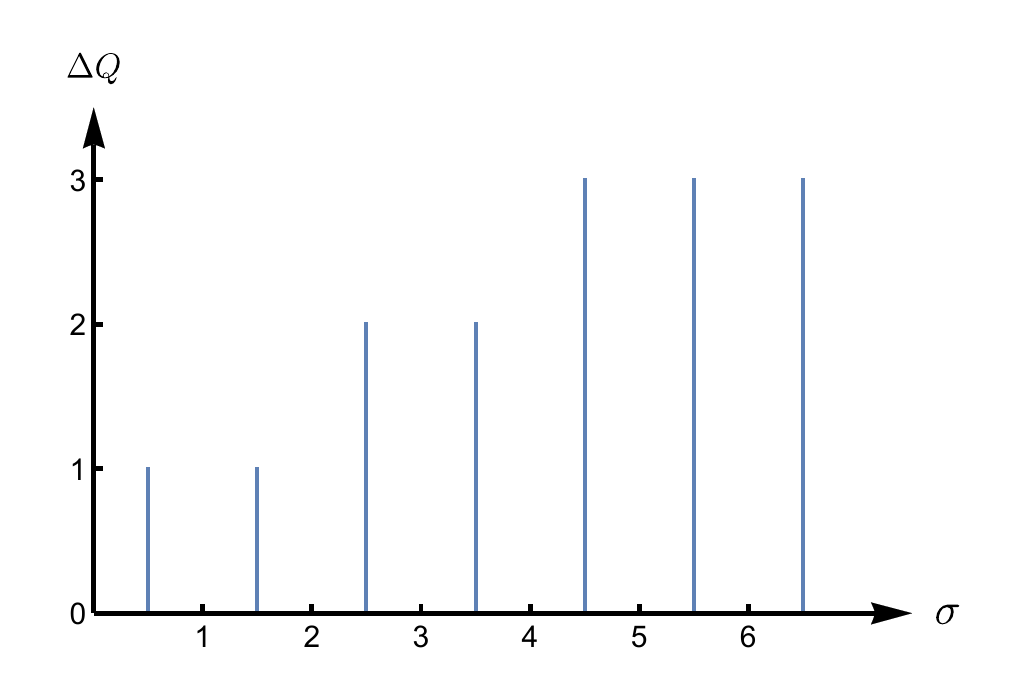} 
\caption{Peaks of tunneling electrons obtained by varying the number
  of flux quanta $\sigma$ (area) of an isolated droplet with $\nu=3$
  (see Eq.(\ref{coulomb})).}
\label{fig5}
\end{figure}

In conclusion, a possible signature of the orbital spin could be
seen at the beginning of the deformation, i.e. for
small $\s$ values. The sequence of electron 
tunnelings would be, for $\nu=n$, 
\be
\D Q=1,1,2,2,3,3,\dots,n,n,n,\dots,\quad {\rm for}\quad
\s+\frac{1}{2}=1,2,\dots,2n,2n+1,2n+2\dots,\qquad (\nu=n),
\label{coulomb}
\ee
leading to a triangular comb plot for $\D Q(\s)$
(see Fig.\ref{fig5}).

%-6.2---------------------------------------

\subsection{Quadrupole deformation}

Another test of the orbital spin at the edge is made
by deforming the shape of the droplet. This is
suggested by the effective action, because the boundary terms (\ref{b-act})
couple $\ov{s}$ and $\ov{s^2}$ to the extrinsic curvature $K$, that
measures shape deformations\footnote{
We thank P. Wiegmann for suggesting this possibility to us.}

Let us consider a quadrupole deformation of the confining potential 
\cite{quadrup},
\be
H\ \to\ H +V_\eps=H+\frac{\eps}{R} \left( z^2+\bar{z}^2 \right),
\ee
where $\eps \ll 1$. We analyze this effect to first perturbative 
order in $\eps$.
Let us go back to the discussion of the confining potential
before the limit to the edge, in appendix B. 
The unperturbed states $\ket{i,m}$ have energy 
$E_{m}^{(i)}=(v/R)(m+2i+1)$, where $i$ and $m$ are level index and momentum.
In the case of $\nu=n$, this unperturbed spectrum is $n$-times degenerate 
for given value of $k=m+2i$.
The expectation value of $V_\eps$ in this degenerate subspace form
a $(n\times n)$ matrix whose eigenvalues give the
the leading perturbative correction to the energy.
The matrix is:
\ba
\left(V_\eps\right)_{ik,jk}
&=&\frac{\eps}{R}\bra{i,m}2b^\dag a+2ba^\dag\ket{j,\ell}
\vert_{m+2i=\ell+2j=k}
\nl
&=& \frac{2\eps}{R}\left(\sqrt{(i+1)(k-i)}\, \d_{i,j-1} +
\sqrt{i(k-i+1)}\, \d_{i,j+1}\right) ,
\nl
&& i,j=0,\dots,n-1 .
\label{matrixel}
\ea
The eigenvalues of this matrix are e.g. $(\pm 2\eps\sqrt{k}/R)$
for two Landau levels and $(0,\pm 2\eps\sqrt{3k-2}/R)$ for $\nu=3$, et
cetera.

Following the discussion of section 4, we should evaluate this correction
in the edge limit, given by 
$R\to\infty$ with $r=R+x$ and $k=m+2i=R^2+m'+2i$. 
We can thus approximate $k-i$ with $R^2$ in the matrix elements 
(\ref{matrixel}), simplifying the eigenvalue problem.
We then find the following modification of the edge spectrum (\ref{spec-m})
(for $b=0$):
\be
v\bra{j,L+m'} 2x+\eps \cos 2\th\ket{j,L+m'}
=\frac{v}{R}\left(m'+2j+1+2\eps \a_j\right).
\label{quadr-spect}
\ee
In this equation, 
$j=0,\dots,n-1$ is the index of the Landau levels (now mixed among themselves 
by the perturbation) and $\a_j$ are $O(1)$ constants, 
whose first few values are:
\be
\begin{array}{l|c|c|c|c}
\hline
n & 2& 3 &4&\dots
\\
\a_j & \pm 1& 0,\pm\sqrt{3}&\pm\sqrt{3\pm\sqrt{6}}&\dots
\\
\hline
\end{array}.
\ee 
In conclusion, the quadrupole perturbation amounts, in the edge
limit, to rigid $O(1/R)$ translations of the branches of the edge spectrum
among themselves. Upon tuning $\eps$, one can exactly compensate the
translation due to the shift, e.g. by setting 
$2j+2\eps \a_j=0$ for a given value of $j$ in (\ref{quadr-spect}).  This
result is consistent with the mentioned coupling of the orbital spin
to the extrinsic curvature.

One physical application of the quadrupole perturbation of the Hall
droplet could be the following. In the Coulomb blockade setting, a
small non-integer shift among the levels could be useful to split the
degeneracy of the peaks. For example, let us consider the value of
$\s$ at which two pairs of levels become degenerate, belonging to two
branches of edge states, causing a $\D Q=2$ peak (see Fig.\ref{fig5}).
In presence of the quadrupole deformation, this double peak splits in
two $\D Q=1$ peaks, occurring at slightly different values of
$\s$. The same pattern repeats itself at distance $\D \s=1$.  This
fact could help interpreting the experimental results.

\section{Conclusions}

In this work we explicitly found the structure of edge
excitations in the multicomponent case, and showed that the various
branches are associated with radial moments of bulk microscopic
states in the edge region.  For integer fillings, 
we did a straightforward analysis of
the large-system limit $R\to\infty$, while in fractional case we
studied the $W_\infty$ deformations of incompressible fluid states.

Using these results, we were able to identify 
Casimir-like ground-state values of charge and conformal spin in the
edge theory that depend on the orbital spin,
in agreement with the effective field theory approach \cite{AGJ}.
This Casimir effect, or chemical potential shift, 
is irrelevant for a single branch of edge excitations, or for several but 
independent edge branches, that may occur in the integer Hall effect.

On the contrary, in the case of several interacting branches, such as
the hierachical Jain states, these ground-state charge and spin are
observable in isolated systems, and are parameterized by the integers
$s_i-s_j$, $i,j=0,1,\dots,n-1$.  We briefly discussed the Coulomb
blockade experiment where the ground-state charge could be measured, while
other signatures of this effect remain to be investigated.

Finally, another line of development concerns the subleading terms in
the $1/R$ expansion discussed in Appendix A, that correspond to
non-relativistic corrections of the conformal theory and may give
other physical effects parameterized by the orbital spin \cite{CR}.

\medskip
{\bf Acknowledgments}

The authors would like to thank A. G. Abanov, D. Bernard, A. Gromov,
T. H. Hansson, K. Jensen,  S. Klevtsov and P. Wiegmann 
for interesting scientific exchanges.
A. C. acknowledge the hospitality and support by the \'Ecole Normale
Sup\'erieure, Paris, and the G. Galilei Institute for
Theoretical Physics, Arcetri, where part of this work was done.

\appendix

%-app A---------------------------------------------------

\section{Higher-order terms in the Moyal brackets}

The second-order term  $O(1/B^2)$ of
the Moyal brackets (\ref{Moyal}) for the variation of the first Landau
level density takes the form: 
\begin{equation}
\delta\wt\rho^{(0)}=\frac{2i}{B^{2}} 
\left(\bar{\partial}^{2}\rho^{(0)}\right)
\partial^{2}w+h.c.\; .
\end{equation}
Remembering that $\rho=\rho^{(0)}\left(r\right)$, we rewrite it as follows:
\begin{equation}
\delta\wt\rho^{(0)}=\frac{2i}{B^{2}}
\left[\left(\frac{\partial}{\partial r^{2}}\right)^{2}\rho^{(0)}\right]
\left(z^{2}\partial^{2}-\bar{z}^{2}\bar{\partial}^{2}\right)w.
\label{sec-ord-Moy}
\end{equation}
The action of the operator,
\begin{equation}
D_{2}=z^{2}\partial^{2}-\bar{z}^{2}\bar{\partial}^{2},
\end{equation}
on a generic polynomial  $w\sim z^{n}\bar{z}^{m}$ is given by:
\begin{equation}
\label{eigen-D2}
D_{2}z^{n}\bar{z}^{m}=\left(n-m\right)\left(n+m-1\right)z^{n}\bar{z}^{m}.
\end{equation}
The second-order correction (\ref{sec-ord-Moy}) to the edge current modes 
(\ref{def-mom}) can be easily computed by integrating the density in 
space and by taking the edge limit (\ref{edge-lim-den}); we find:
\be
\d\wt\r^{(0)}_k=\frac{1}{R}\int dx \; x e^{-2x^2}\; iD_2 w(x,\th),
\qquad\quad (B=2).
\label{1LL-fluct2}
\ee

This expression is different from the leading contribution 
$\d\r^{(0)}_k$ in Eq.(\ref{1LL-fluct}) for 
two reasons. The first is the order $O(1/R)$ that is subleading
w.r.t. the $O(1)$ behaviour entering the conformal current algebra 
(\ref{chi-algebra}).
The second aspect is that the dependence on the Fourier 
modes (\ref{1LL-fluct2}) given by the
operator $D_2$  does not appear in the conformal operators $\r_k$
and $L_\k$ discussed earlier (cf. (\ref{viras}), (\ref{modes})).
Actually, this spectrum is associated to a higher-spin conserved current of 
the conformal theory,
the dimension-three current $\hat V^2(\eta)$, whose expression in
the fermionic conformal theory is \cite{W-dyna}
\be
\hat V^2_0=\frac{1}{R^2}\sum_r \left(r-\frac{1}{2}\right)^2
:\hat d^\dag_r \hat d_r:\; ,
\label{high-spin}
\ee
(note the quadratic weight in the sum w.r.t. the linear one of 
$\hat L_0$ (\ref{viras})).
Actually, the action of $\hat{V}^2_0$ on a particle-hole excitation, 
$\hat d^\dag_n \hat d_m\ket{\W}$ is readily found to reproduce the
eigenvalue of $D_2$ in (\ref{eigen-D2}).
The operator $\hat{V}^2_0$ naturally appears as a subleading $1/R^2$
term in the Hamiltonian (\ref{confo-ham}) of the conformal theory on
the cylinder and corresponds to a non-relativistic correction to the
edge dynamics \cite{W-dyna}.  In summary, the second order term in the
Moyal brackets represents a non-relativistic corrections to the
conformal theory of edge excitations.

We remark that such a correction is expected to express another
physical effect of the orbital spin $s$.  The work \cite{CR} presented
an analysis of the $1/B$ correction to the effective theory of section
2: it involves a spin-two hydrodynamic field in the bulk 
$b_k=b_{\mu k}dx^\mu$, where $k$ is a space index, and a generalized action with
coupling constant proportional to $s$.  In analogy with the discussion
in section 2, this action requires a boundary term that involves the
edge theory: the corresponding contribution to $\hat{H}$ is expected
to be given by (\ref{high-spin}).

In conclusion, the $O(1/B^2)$ term in the Moyal brackets describes
a non-relativistic corrections to the conformal theory that is
parameterized by the orbital spin. The complete analysis of this effect
is left to future developments of this work.
Let us mention the related approach \cite{moroz} discussing the
non-relativistic correction to the Hall conductivity due to the orbital
spin.

A final remark concerns the relation between the $1/B$ series
in the bulk and the $1/R$  expansion at the edge defined in section
3. To leading order in $1/B$, a finite limit is found
for bulk quantities in the rescaled coordinate $u$, 
with $r=Ru\sim \sqrt{N} u$.
For example, the limit of the ground-state density of Laughlin
states is given by the step function, 
\be
\r(\sqrt{N}u)=\frac{\nu B}{2\pi}\Theta(1-u). 
\ee
In this limit, the magnetic length is sent to zero, owing to $\ell^2=2/B$, 
and the edge region shrinks to a point.
On the contrary, in the edge limit $R\to \infty$ the relevant features
take place in the shell $x=r-R=O(\ell)$, where the wavefunctions have
support and the Hermite polynomial expansion takes place. This region
should be kept finite.
Therefore, the $1/B$ approximation
is not accurate enough for the analysis of edge excitations because it is too
singular in that region.

In conclusion, we identified the first subleading term that is
expressed by the spin-three conserved current \cite{W-dyna} and
parameterized by the orbital spin.  Being observable in the conformal
theory, such correction could give physical sense to $s$ also for
single-branch excitations.  The bulk-boundary correspondence for
subleading corrections will involve the study of the effective theory
introduced in Ref.\cite{CR} within the multipole expansion of bulk
excitations.

%-app B------------------------------------------------------

\section{Confining potentials}

The simplest model is obtained by adding a quadratic confining potential 
$V=v|z|^2/R$ to the Landau level Hamiltonian in (\ref{landau-ham}), as follows:
\be
H^{(2)}= H+ V=
2 a^\dag a +1 +\frac{v}{R}\left(a^\dag a +b^\dag b+1 +a^\dag b^\dag+
a b \right),
\label{conf-ham}
\ee
where $z$ and $\bar z$ are expressed in terms of the $a$ and $b$ 
oscillators (\ref{ab-oper}). 
The exact spectrum of $H$ (\ref{conf-ham}) can be
obtained by a Bogoliubov transformation in the two-dimensional space
of the oscillator pair $(a,b)$, leading to the new
pair $(A,B)$ defined by:
\begin{empheq}[left=\empheqlbrace]{align}
\label{bog1}
  A    &= \cosh\left(\f_{b}\right)a+\sinh\left(\f_{b}\right)b^{\dagger},\\
  B    &= \cosh\left(\f_{b}\right)b+\sinh\left(\f_{b}\right)a^{\dagger},
\label{bog2}
\end{empheq}
with $\tanh(2\f_{b})=\frac{v}{v+R}$.
The result is:
\be
H^{(2)}=\left(1+\sqrt{1+\frac{2v}{R}}\right) A^\dag A+
\left(\sqrt{1+\frac{2v}{R}}-1\right) B^\dag B +{\rm const.}\;.
\label{bog-ham}
\ee 
After the transformation, the angular momentum is $J=B^\dag
B-A^\dag A$, with eigenvalue $m$, and the $A^\dag A$ eigenvalue $n$
gives the dressed Landau levels energies.

We now obtain the energy spectrum of the edge excitations by
evaluating the spectrum of \eqref{bog-ham} in the limit to the edge
defined in section three, namely $|z|=r=R+x$, $R\to\infty$ with
$m=L+m'$, $L=R^2$, $|m'|<\sqrt{L}$.
The result is:
\be 
\e_{n,m'}^{(2)}=2n+vR+v\frac{m'+2n+1}{R}+O\left(\frac{m'^2}{R^2}\right)+
\text{const.},
\label{ed-spec2}
\ee
where the divergent term must be subtracted from the potential,
$V\rightarrow V-vR=2x+x^2 /R$, to obtain a finite linear term.

The spectrum with other potentials $V\sim r^k/R^{k-1}$ is necessary to
disentangle the contribution of the linear and quadratic terms in the
spectrum $\e^{(2)}_{n,m}$ and to evaluate the higher terms $x^3 /R^2$,
etc\ldots .  The Bogoliubov transformation does not apply to generic
potentials, and we use another method that works in the relevant limit
$R\to\infty$. Note that the angular momentum $J=b^\dag b- a^\dag a$ is
a good quantum number for any potential and it takes very large values
in this limit; this means that the expectation values of $b^\dag b$ is
of order $R^2$. Remembering the case of Bose-Einstein condensation, we
can treat the $b$ operator as a classical variable, and write:
\be
b\sim b^\dag \sim\sqrt{b^{\dagger}b}=\sqrt{R^2+m'+a^{\dagger}a}
\simeq R\left(1+\frac{m'+a^{\dagger}a}{2R^{2}}\right)\sim R+
O\left(\frac{1}{R}\right).
\label{b-app}
\ee
We use this approximation in the original Hamiltonian \eqref{conf-ham} and 
find the expression:
\be 
H^{(2)}=2\left(1+\frac{v}{R}\right)a^{\dag}a+v\frac{m'}{R}+
v\left(a^{\dag}+a\right)+\text{const.}.
\label{ed-ham2}
\ee

This result can be checked with the $R\to\infty$ limit of the
Bogoliubov transformation (\ref{bog1}, \ref{bog2}). Note that the
rotation angle $\f_{b}$ is small:
\begin{empheq}[left=\empheqlbrace]{align}
\label{bog-bigR1}  
  A    &\simeq a+\frac{v}{2R}b^\dag , \\
  B    &\simeq b+\frac{v}{2R}a^{\dag}.
\label{bog-bigR2}
\end{empheq}
Substituting these expressions and the approximation \eqref{b-app}
into the exact result \eqref{bog-ham}, we reobtain the expression
\eqref{ed-ham2}.  Furthermore, the result \eqref{ed-ham2} also agrees
with the $R\to\infty$ limit of the exact matrix elements of the
quadratic potential,
\ba
&&\bra{i,R^2+m'}V^{(2)}\ket{j,R^2+m'}\simeq
\nl
&&\qquad \qquad\quad
v\left[\left(R+ \frac{m'+2i+1}{R}\right)\d_{i,j}+
\sqrt{j+1}\d_{i,j+1}+\sqrt{j}\d_{i,j-1}\right],
\label{me-quad}
\ea
that can be evaluated using the wave functions 
\eqref{Landau-wfunct} and the help of Mathematica.

We now diagonalize the approximate $H^{(2)}$ \eqref{ed-ham2} directly. 
We note that a shift of the operator $a$ by a constant, setting
\be 
A=a+\frac{v}{2}\left(1-\frac{v}{R}\right),
\ee
reproduces the $R\to\infty$ limit of the Bogoliubov 
transformation \eqref{bog-ham}. 
In conclusion, the spectrum \eqref{ed-spec2} has been recovered 
by using directly the approximation \eqref{b-app} in the Hamiltonian $H^{(2)}$.

We extend the previous analysis to the cases of linear and quartic
confining potentials, $V^{(1)}=v_{1}r$ and
$V^{(4)}=v_{4}r^{4}/R^3$. In the linear case, the Hamiltonian is:
\be 
H^{(1)}=2a^{\dagger}a+1+\frac{v_{1}}{R}
\left(a^{\dagger}a+b^{\dagger}b+1+a^{\dagger}b^{\dagger}+ab\right)^{1/2}.
\ee
We apply the large $R$ limit and the approximation \eqref{b-app}, 
subtract the infinite term and find:
\be 
H^{(1)}=2a^\dagger a+\frac{v_{1}}{2}\left(a^{\dagger}+a\right)+
\frac{v_{1}}{8R}\left(6a^\dagger a +4m'\right)-
\frac{v_{1}}{8R}\left(a^{\dagger 2}+a^2\right)+\text{const.}.
\label{ed-ham1}
\ee
The calculation of the matrix elements of $V^{(1)}$
in the edge limit agrees with the previous result:
\ba 
&&\bra{i,R^2+m'}V^{(1)}\ket{j,R^2+m'}\simeq
\nl
&&\qquad\qquad \qquad 
v_{1}\left[\left(\frac{6i+4m'}{8R}+R\right)\delta_{i,j}+
\frac{v_{1}}{2}\left(\sqrt{i-1}\delta_{i-1,j}+\sqrt{i}\d_{i,j-1}\right)\right.
\nl
&&\qquad\qquad \qquad 
\left. +\frac{v_{1}}{8R}\left(\sqrt{(j+2)(j+1)}\d_{i,j+2}+
\sqrt{j(j-1)}\d_{i,j-2}\right)\right].
\label{me-lin}
\ea
The equation \eqref{ed-ham1} involves an additional term
$\left(a^{\dag 2}+a^2\right)$ w.r.t the quadratic case \eqref{ed-ham2}
corresponding to non-vanishing matrix elements on the third
diagonal. After the shift by constant of the operator $a$, the
Hamiltonian $H^{(1)}\sim A^\dag A+\l \left(A^{\dag 2}+A^{2}\right)$
can be diagonalized by a change of the oscillator frequency which
turns out to be $O(1/R^2)$.
The diagonal form of the Hamiltonian $H^{(1)}$ is finally,
\be 
H^{(1)}\simeq \left(2+\frac{3v_{1}}{4R}\right)A^{\dagger}A+
v_{1}\frac{m'}{2R}+\text{const.},
\label{hamd-lin}
\ee
with spectrum:
\be
\e^{(1)}_{n,m'}=2n+v_{1}\left(\frac{m'}{2R}+\frac{3n}{4R}\right)+\text{const.}.
\label{ed-spec1}
\ee

In the case of quartic potential, the approximated Hamiltonian 
can be written as follows:
\be 
H^{(4)}=2a^{\dagger}a+1+\frac{v_{4}}{R}\left(6a^{\dagger}a+2m'\right)+
2v_{4}\left(a^{\dagger}+a\right)+\frac{v_{4}}{R}\left(a^{\dagger 2}+
a^2\right)+\text{const.},
\label{ham-4}
\ee
in agreement with the matrix elements,
\ba
&&
\bra{i,R^2+m'}V^{(4)}\ket{j,R^2+m'}\simeq 
\nl
&&\qquad\qquad
 v_{4}\left[\frac{1}{R}\left(6i+2m'+\text{const.}\right)\d_{i,j}+
2\sqrt{j+1}\d_{i,j+1}+ 2\sqrt{j}\d_{i,j-1}\right. 
\nl
&&\qquad\qquad
\left.+\frac{1}{R}\left(\sqrt{\left(j+1\right)
\left(j+2\right)}\d_{i,j+2}+\sqrt{j\left(j-1\right)}\d_{i,j-2}\right)\right].
\label{me-quar}
\ea
We can diagonalized \eqref{ham-4} by introducing once again a constant shift in operator $a$ and the frequency redefinition, leading to the spectrum:
\be 
\e^{(4)}_{n,m'}=2n +v_{4}\left(\frac{2m'}{R}+\frac{6n}{R}\right)+\text{const.}.
\ee

Now let us discuss the generic confining potential \eqref{spec-x} for $r=R+x\to\infty$:
\be 
V(x;R)=a_{1} x+\frac{a_{2} }{R}x^{2}+\frac{a_3}{R^{2}}x^{3}+\ldots .
\label{ed-con-pot}
\ee 
We are interested in the relativistic conformal spectrum at the edge \eqref{lin-spectrum}, which is of order $O(1/R)$. We compare the spectrum of $V^{(2)}$, $V^{(1)}$ and $V^{(4)}$ computed in this limit and obtain the following consistent result for the spectrum:
\be
\e_{n,m'}=a_1\left(\frac{m'}{2R}+\frac{3n}{4R}\right)+a_{2}\frac{n}{2R}+\text{const.}.
\ee
This comparison shows that the higher terms $x^3 /R^2$ that appear for example in $V^{(4)}$, do not contribute to leading order $O(1/R)$. This fact can also be checked by evaluating the corresponding matrix elements in the edge limit \eqref{nLL-edge}.

In conclusion, the relativistic conformal spectrum takes values from the linear and quadratic terms in \eqref{ed-con-pot}.

%---------------------------------------------

\end{document}